\documentclass[a4paper,fleqn,usenatbib]{mnras}{\normalsize }

\usepackage[T1]{fontenc}
\usepackage{ae,aecompl}

\usepackage{graphicx}   
\usepackage{amsmath}    
\usepackage{amssymb}    
\usepackage[dvipsnames]{xcolor}
\newcommand{\etal}{et al.}

\def\lesssim{\mathrel{\hbox{\rlap{\hbox{\lower4pt\hbox{$\sim$}}}\hbox{$<$}}}}
\def\gtrsim{\mathrel{\hbox{\rlap{\hbox{\lower4pt\hbox{$\sim$}}}\hbox{$>$}}}}
\def\apj{ApJ}
\def\apjs{ApJS}
\def\aj{AJ}
\def\aap{A\&\hskip-1pt A}
\def\aaps{A\&\hskip-1pt AS}
\def\mnras{MNRAS}
\def\pasp{PASP}

\def\jkas{JKAS}

\title[SFHs of Early-type Dwarfs]{Star Formation Histories of Early-type Dwarfs in Group Environment}

\author[Ann et al.]{Hong Bae Ann$^1$, Mira Seo$^1$ Myoungwon Jeon$^{2,3}$\thanks{E-mail:
myjeon@khu.ac.kr}\\ \\ 
$^{1}$Department of Earth Science Education, Pusan National University, 46241, Busan, Republic of Korea\\
$^{2}$School of Space Research, Kyung Hee University, 1732 Deogyeong-daero, Yongin-si, Gyeonggi-do 17104, Korea\\
$^{3}$Department of Astronomy \& Space Science, Kyung Hee University, 1732 Deogyeong-daero, Yongin-si, Gyeonggi-do 17104, Korea
}

\date{Accepted XXX. Received YYY; in original form ZZZ}

\pubyear{2022}

\begin{document}
\label{firstpage}
\pagerange{\pageref{firstpage}--\pageref{lastpage}}
\maketitle



\begin{abstract}
	
	We investigate the star formation histories (SFHs) of 983 early-type dwarf galaxies classified into five morphological subtypes—dS0, dE, dE$_{\mathbf{bc}}$, dSph, and dE$_{\mathbf{bl}}$—across six environments ranging from the field to rich clusters such as Ursa Major and Virgo. Using full spectral fitting of SDSS spectra with the \textsc{starlight} code, we derive detailed SFHs and chemical enrichment patterns. We find that SFHs are primarily shaped by morphology, with environment playing a secondary but non-negligible role. Red early-type dwarfs (dS0, dE, dSph) typically formed most of their stars early and quenched rapidly, whereas blue early-type dwarfs (dE$_{\mathbf{bc}}$, dE$_{\mathbf{bl}}$) exhibit extended or ongoing star formation and host extremely metal-poor stars, suggesting continued pristine gas accretion. Environmental dependence is clearest in low-mass systems: field galaxies often show prolonged SFHs and delayed enrichment, while Virgo Cluster galaxies tend to quench earlier and enrich more rapidly. Cumulative SFHs reinforce these trends, with dSph galaxies showing the earliest quenching and least environmental dependence, indicating a likely primordial origin. Metallicity evolution also varies with mass and environment, progressing most slowly in low-mass field galaxies and most rapidly in high-mass cluster galaxies. Our results highlight the combined influence of morphology, stellar mass, and environment on the evolutionary diversity of early-type dwarfs, and suggest that both internal processes (nature) and external conditions (nurture) are intricately linked in shaping their star formation and chemical enrichment histories.
	
\end{abstract}

\begin{keywords}
 galaxies: dwarfs -- galaxies: formation -- galaxies: evolution -- galaxies: star formation 
\end{keywords}

\section{Introduction}

Dwarf galaxies are the most numerous type of galaxy in the Universe, dominating both low-density field environments and denser group and cluster environments. Among them, early-type dwarfs are predominantly found in groups and clusters, typically as satellites of more massive galaxies, whereas late-type dwarfs are more commonly observed in the field. Interestingly, a subset of early-type dwarfs also resides in the field. These field early-type dwarfs are considered genuine primordial systems, as they are less likely to have undergone the frequent gravitational interactions that drive morphological transformation in denser environments. As such, they serve as valuable laboratories for probing the star formation histories (SFHs) of galaxies during the early stages of cosmic evolution.

Most previous investigations of the SFHs of early-type galaxies have relied on color-magnitude diagrams (CMDs) constructed from resolved stellar populations down to the main-sequence turn-off. Due to the low luminosity of such stars, CMD-based analyses have been largely limited to galaxies within the Local Group (LG) \citep{her00, apa01, dol02, car02, lee09, mon10a, mon10b, boe12, hid12, hid13, bro14, wei14, gal15, sav15, san16, mak17, ski17, bet18, bet19, sav19, wei19, gal21, rus21, nav21}, with only a few exceptions \citep[e.g.,][and references therein]{wei11, cig19}. 

Beyond the LG, alternative approaches such as full spectral fitting have been employed to study unresolved stellar populations in dwarf galaxies. Notably, this method has been applied to 12 dSph galaxies in the Centaurus A (Cen A) group \citep{mul21} and to dwarf galaxies in the Fornax Cluster \citep{rom23, rom24a, rom24b}, using integrated spectra obtained with the MUSE integral field spectrograph on VLT UT4 \citep{bac10} and the SAMI instrument on the 3.9\,m Anglo-Australian Telescope \citep{cro12}, respectively. The dSphs in the Cen A group are characterisedby old, metal-poor stellar populations and follow the stellar metallicity–luminosity relation observed in LG dwarf galaxies. In contrast, the SFHs of dwarfs in the Fornax Cluster exhibit strong environmental dependence, with early quenching observed in galaxies with stellar masses below $10^8\,M_{\odot}$.

Recently, the SFHs of early-type dwarf galaxies have been investigated through analysis of optical spectra using population synthesis techniques \citep{seo23, ann24}. These studies revealed that the majority of early-type dwarfs, including dS0s, dEs, SFHs typically show two distinct periods of star formation, with peaks occurring approximately 10 Gyr and 2.5 Gyr ago. However, the strength of the second peak at 2.5 Gyr varies significantly among morphological subtypes: dS0s and dEs display secondary peaks nearly as prominent as the initial burst, while dSphs show a much weaker secondary episode. The most striking result that was observed in the SFHs of the early-type dwarfs is that the morphology distinction given by their subtypes is most important for distinguishing the properties of SFHs. 

Additional subtypes of early-type dwarf galaxies have been identified by \citet{ann15}. One such subtype is the blue-cored dwarf elliptical galaxy, denoted as dE$_{\mathbf{bc}}$ and the other is the globally blue dwarf elliptical or spheroidal galaxy, classified as dE$_{\mathbf{bl}}$. 
The dE$_{\mathbf{bc}}$ subtype has been widely reported in the literature \citep{vig84, bin91, pel93, dur97, lot04, lis06, lis07, gu06, chi08, tul08, ge12, kim14, pak14, ann15, uri17, chu19, chu21, rey23}, with early examples including NGC 185 \citep{hod63} and NGC 205 \citep{hod73}. In contrast, the dE$_{\mathbf{bl}}$ subtype was introduced more recently by \citet{ann15} and has not yet been extensively studied. These galaxies are considered transitional systems, sometimes referred to as dE/I types \citep{tul08} and they resemble blue compact dwarfs (BCDs) in morphology and star-forming properties. The colors of dE$_{\mathbf{bl}}$ galaxies are comparable to those of dwarf irregulars (dIs), although typically less blue than BCDs, suggesting the presence of ongoing star formation, albeit at a lower intensity than in BCDs. Notably, about half of the dE$_{\mathbf{bc}}$ population and the majority of dE$_{\mathbf{bl}}$ galaxies are found in low-density environments such as the field \citep{seo22}.

Since most dS0, dE and dSph galaxies are located in group and cluster environments, it is of particular interest to examine the environmental dependence of the cumulative star formation histories (cSFHs) of early-type dwarfs, including the dE$_{\mathbf{bc}}$ and dE$_{\mathbf{bl}}$ subtypes. In this study, we consider environments spanning the field, poor groups, rich groups and clusters. To this end, we identify groups and clusters in the local universe using galaxies from the Catalog of Visually Classified Galaxies at $z \lesssim 0.01$ \citep[][hereafter CVCG]{ann15}. It should be noted that the dwarf galaxies in our sample are relatively luminous, as the CVCG is constructed from the Sloan Digital Sky Survey \citep[][hereafter SDSS]{yor00}, which has a spectroscopic limiting magnitude of $r = 17.77$. This corresponds to $M_r = -13.4$ at the distance of the Virgo Cluster and $M_r = -15.2$ at the redshift limit of the CVCG. Despite this luminosity bias, the influence of environment on the cSFHs of early-type dwarfs can still be meaningfully assessed. Previous studies have demonstrated that the local background density traced by such galaxies significantly affects the SFHs of dSphs and dEs \citep{seo23}, as well as dS0s \citep{ann24}.

Among early-type dwarf galaxies, dSphs are closely associated with early quenching, whereas dE$_{\mathbf{bc}}$ and dE$_{\mathbf{bl}}$ galaxies are believed to be linked to late accretion processes. Therefore, examining the SFHs of these subtypes is essential for understanding galaxy formation across cosmic time. Some of these galaxies are considered primordial, providing insights into the early stages of galaxy formation, while others are thought to have formed relatively recently, illustrating the downsizing phenomenon. At the same time, the well-established morphology–density relation \citep{dre80} implies that morphological properties are inherently connected to the environment. Thus, interpreting the SFHs of early-type dwarfs requires a consideration of their environmental context. In this study, we investigate the dependence of cSFHs on environment, focusing on group properties that span field to cluster environments.

The structure of this paper is as follows. In Section~2, we describe the selection of the sample galaxies and provide a brief overview of the SDSS spectral data. Section~3 outlines the methodology used to derive the star formation histories of early-type dwarfs. The cumulative star formation histories are presented in Section~4, followed by a discussion in Section~5. Our conclusions are summarized in Section~6.

\section{Data and Method}
\subsection{Data}

We utilized spectroscopic data from the Sloan Digital Sky Survey (SDSS) for a sample of early-type dwarf galaxies whose morphological classifications were provided by \citet{ann15}. Among the 1,089 classified galaxies, 983 were retained for analysis, as the remaining objects lacked spectra of sufficient quality. The early-type dwarf galaxies in the CVCG are categorized into five subtypes: dwarf lenticulars (dS0), dwarf ellipticals (dE), blue-cored dwarf ellipticals (dE$_{\mathbf{bc}}$), dwarf spheroidals (dSph) and blue dwarf ellipticals (dE$_{\mathbf{bl}}$). A comprehensive description of their morphological characteristics is available in \citet{ann15}. The CVCG is nearly complete for galaxies brighter than $r = 17.77$ within the SDSS survey footprint. The final sample used in this study includes 148 dS0s, 234 dEs, 183 dE$_{\mathbf{bc}}$s, 200 dSphs and 218 dE$_{\mathbf{bl}}$s. The early-type dwarfs in the CVCG are not defined by luminosity but by morphology. However, as shown in the figure, our sample of dwarf galaxies has $M_{r} \gtrsim -18$.

\begin{figure}
	\centering
	\includegraphics[width=\columnwidth]{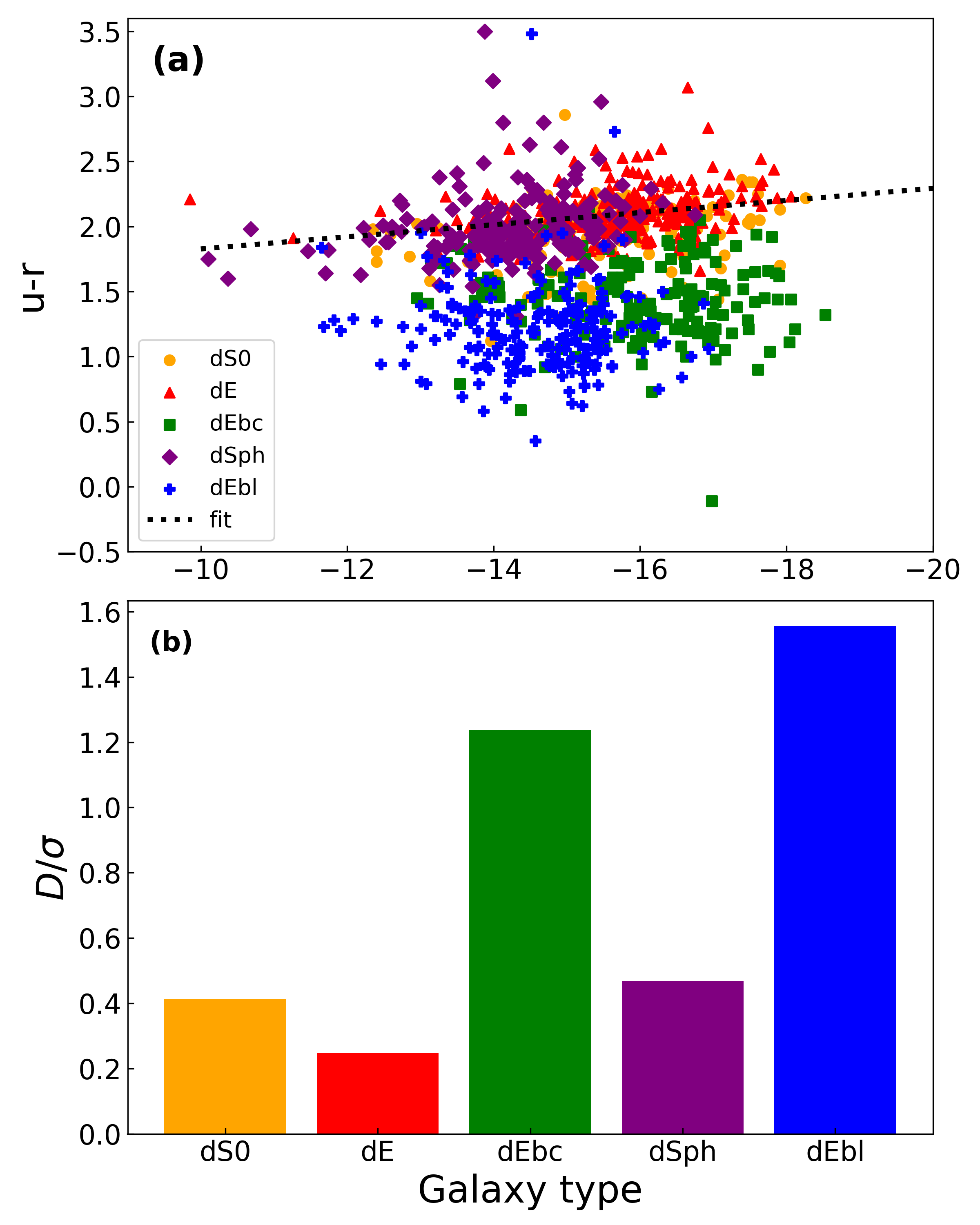}
	\caption{a) Colour--magnitude diagram for early-type dwarf galaxies. The five subtypes are distinguished by symbols and colours as indicated in the legend. The dotted black line shows the linear fit to the combined distribution of dS0, dE, and dSph galaxies, which defines the fiducial red sequence.  
		(b) Mean deviation from the red sequence, normalised by the RMS dispersion ($D/\sigma$), for each morphological subtype.}
	\label{fig1}
\end{figure}

Figure~\ref{fig1}-(b) shows the mean deviations from the fitted red sequence, normalized by the RMS dispersion of galaxies that define the red sequence (i.e. dS0, dE, and dSph), for each subtype. The Y-axis represents this normalized offset, denoted as $D/\sigma$, where $D$ is the average vertical deviation (i.e. $u-r$) from the red sequence and $\sigma$ is the root mean square (RMS) of residuals for the red sequence galaxies. As expected, dS0, dE, and dSph galaxies show relatively small deviations from the red sequence, with the smallest deviation $D/\sigma$ in dE galaxies. In contrast, dE$_{\bold{bc}}$ galaxies exhibit a larger deviation, occupying the so-called green valley. The dE$_{\bold{bl}}$ galaxies show the greatest offset, with a $D/\sigma$ value exceeding 1.5, indicating that they lie well outside the typical colour range of the red sequence. This trend reinforces the interpretation that dE$_{\bold{bc}}$ and dE$_{\bold{bl}}$ galaxies represent transitional or actively evolving populations in terms of their stellar populations and star formation histories.

We used SDSS spectra for the selected early-type dwarf galaxies, retrieved from Data Release 7 (DR7) of the Sloan Digital Sky Survey. The spectra were obtained using $3^{\prime\prime}$-diameter fibers mounted in the focal plane of the 2.5-meter telescope at Apache Point Observatory. The SDSS spectrograph employs 320 fibers, with typical exposure times of 45 minutes or longer to achieve a fiducial signal-to-noise ratio. The spectral coverage spans 3800–9200\,\AA\ with a mean resolving power of $\lambda/\Delta \lambda \sim 1800$. We used fully reduced spectra, calibrated in both wavelength and flux. Since the fiber diameter is significantly smaller than the angular extent of the sample galaxies, the resulting spectra primarily reflect the stellar populations in their central regions. Additional observational parameters, such as distance, $r$-band absolute magnitude (M$_r$), $u-r$ color, morphological type, coordinates, and redshift, are adopted from \citet{ann15}.

\subsection{Methods}

\subsubsection{Population Synthesis}

We applied the STARLIGHT spectral synthesis code \citep{cidF05} to the SDSS spectra of 983 early-type dwarf galaxies to derive their star formation histories. STARLIGHT estimates the most probable combination of stellar populations contributing to the observed spectrum as a function of age and metallicity. The methodology and implementation of STARLIGHT are detailed in \citet{cidF04, cidF05}, with recent applications presented in \citet{rif21} and \citet{seo23}. The code fits each observed spectrum with a model constructed from linear combinations of simple stellar population (SSP) templates drawn from the models of \citet{bc03}.

In STARLIGHT, the model spectrum $M_{\lambda}$ is expressed as

\begin{equation}
  M_{\lambda}=M_{\lambda_{0}}(\sum_{j=1}^{N_{\ast}} x_{j} b_{j, \lambda} r_{\lambda}) \otimes G(v_{\ast},\sigma_{\ast})
\end{equation}

\noindent{where $b_{j, \lambda}$} is the $j-$th SSP spectrum normalized at $\lambda_{0}$,
$r_{\lambda}=10^{-0.4(A_{\lambda}-A_{\lambda_{0}})}$, $M_{\lambda_{0}}$ is
the synthetic flux at the normalization wavelength $\lambda_{0}$ and $x_{j}$ is
the fractional contribution of the SSP for $j-$th population that has age $t_{j}$ and metallicity Z$_{j}$. The stellar motion projected on the line-of-sight is modeled by a Gaussian distribution (G) centered on the galaxy radial velocity $v$ with velocity dispersion $\sigma$. Extinction due to foreground dust is taken into account using the V-band extinction A$_{V}$. The best fitting model is determined by
selecting a model that minimizes the $\Xi^{2}=\Sigma[(O_{\lambda}-M_{\lambda})w_{\lambda}]^{2}$ where $O_{\lambda}$ is the observed spectrum and $w_{\lambda}$ is the inverse of the error applied (see \citet{cidF04} for a detailed description).

We resampled the observed spectra to a uniform wavelength step of $\delta\lambda = 1$\,\AA, after correcting for interstellar reddening and redshift following the procedure described in \citet{seo23}. In the STARLIGHT analysis, each stellar population is characterized by its luminosity fraction ($x_j$) and mass fraction ($\mu_j$), distributed across six metallicity bins (Z = 0.0001, 0.0004, 0.004, 0.008, 0.02 and 0.05), assuming [$\alpha$/Fe] = 0. The SSP models adopted by \citet{bc03} assume a solar metallicity of Z$_{\odot} = 0.02$.

STARLIGHT provides two types of mass fraction for each SSP component: $\mu_{\mathrm{ini}}$, the initial stellar mass, and $\mu_{\mathrm{cor}}$, the mass corrected for gas returned to the interstellar medium. In this study, we used $\mu_{\mathrm{cor}}$ to represent the present-day mass distribution of stellar populations and $\mu_{\mathrm{ini}}$ to estimate star formation rates.

\begin{figure}
	\centering
	\includegraphics[width=0.49\textwidth]{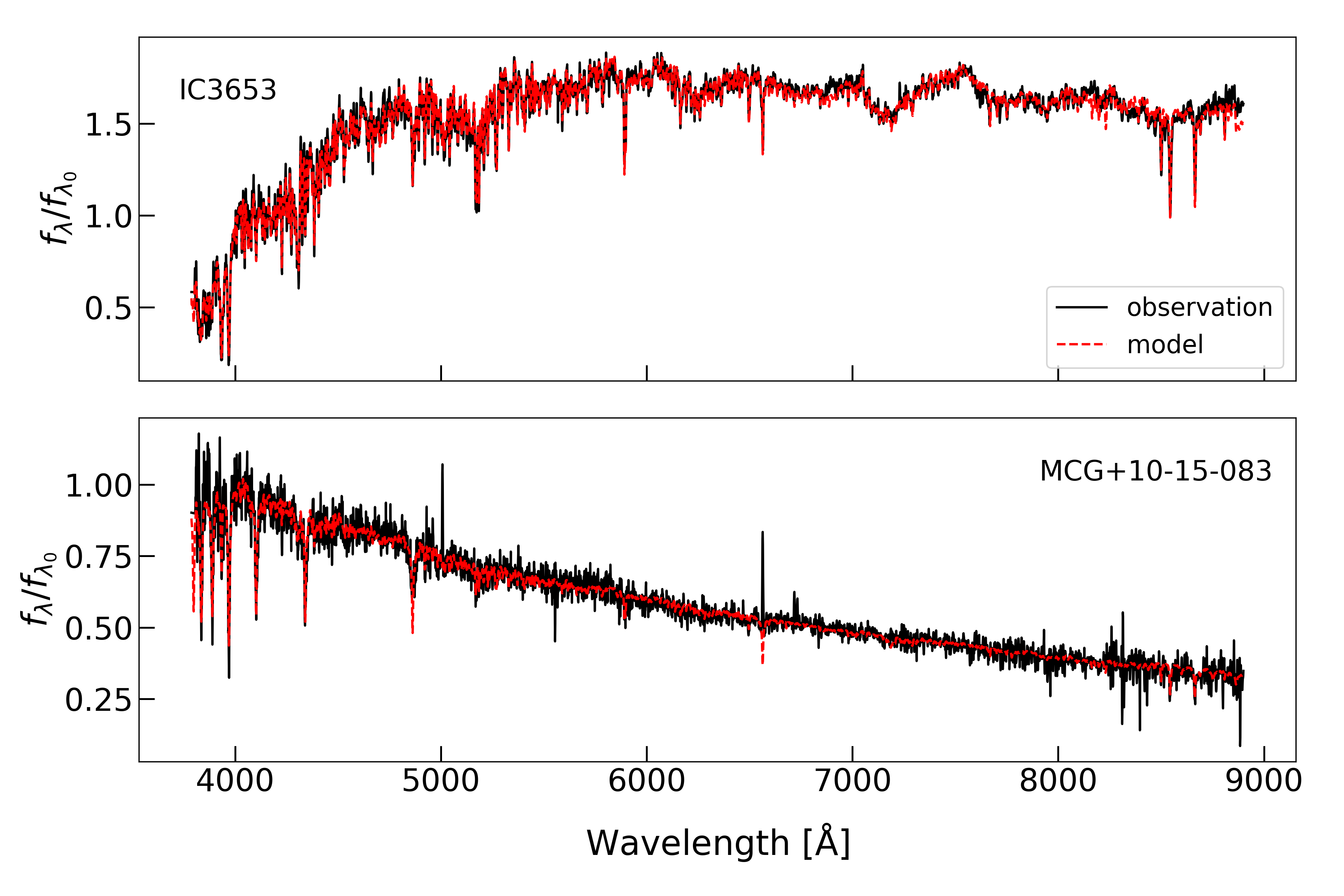}
	\caption{Spectrum of two sample galaxies IC 3653 and MCG+10-15-083. We plot observed spectra  in black and the model spectra in red. Observed and model fluxes are normalized by the flux at $\lambda_0 = 4020\,\text{\AA}$.}
	
	\label{fig2}
\end{figure}

As examples, Figure~\ref{fig2} shows the observed and modeled spectra for two representative galaxies: IC~2653 (a dE galaxy) and MCG+10-15-083 (a dE$_{\mathbf{bc}}$ galaxy). The model spectra provide excellent fits to the observed data. Additional examples of spectral fitting can be found in \citet{seo23} and \citet{ann24}.

The reliability of stellar population properties derived using STARLIGHT has been evaluated in several previous studies \citep{mag15, cidF18, seo23, ann24}. \citet{seo23} tested the robustness of the results using mock spectra generated by perturbing model fluxes with Gaussian noise scaled inversely with the signal-to-noise ratio (S/N). They found that STARLIGHT reliably recovers stellar ages and metallicities for spectra with $S/N \gtrsim 5$. A comparison between STARLIGHT and the pPXF method \citep{cap04, cap17}, conducted by \citet{ann24}, showed good agreement between the two approaches, with a root-mean-square scatter of $\sim$0.3 dex in both mean stellar ages and metallicities. However, a systematic offset was noted: STARLIGHT yields metallicities that are, on average, 0.3 dex higher than those derived using pPXF. A similar discrepancy was reported by \citet{men16} for the dE galaxy NGC~1396 in the Fornax Cluster.

Stellar masses were derived from the model fluxes obtained through STARLIGHT, using galaxy distances listed in the CVCG. An aperture correction was applied following the method described in \citet{ann24}. The resulting stellar masses for the full sample of 983 early-type dwarf galaxies span the range $10^6$–$5 \times 10^9$\,M$_{\odot}$, consistent with the typical mass range of early-type dwarfs.

\subsubsection{Group Finding}

We identified galaxy groups using a friends-of-friends (FoF) algorithm based on the method of \citet{huc82}. The algorithm searches for galaxies that are physically associated with a given target galaxy and if such neighbours are found, the process is recursively repeated for each of them. Two criteria are used to define a physical association: (1) a maximum velocity difference ($\Delta V$) to constrain the relative recession velocity and (2) a maximum projected separation on the sky. The velocity threshold is commonly set to either 500 or 1000\,km\,s$^{-1}$ \citep[e.g.,][]{ann14}. For the spatial linking criterion, we adopt a variable linking length defined by the sum of the virial radii of the target and neighbouring galaxies. That is, two galaxies are considered physically connected if their projected separation is less than the sum of their virial radii.

In this study, we used $\Delta V = 500$\,km\,s$^{-1}$, which is a reasonable choice given the typical peculiar velocities of galaxies \citep{pee79}. Using a larger velocity threshold, such as $\Delta V = 1000$\,km\,s$^{-1}$, results in fewer identified clusters, as more galaxies are merged into larger systems. For example, in the case of the Virgo Cluster, the number of member galaxies increases from 1068 to 1750 when $\Delta V$ is changed from 500 to 1000\,km\,s$^{-1}$, while the total number of identified galaxy groups decreases from 510 to 490.

\begin{figure}
	\centering
	\includegraphics[width=0.45\textwidth]{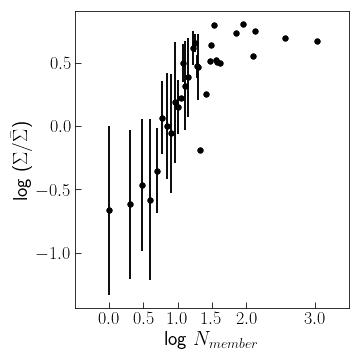}
	\caption{Background density of groups and clusters as a function of the number of group members ($N_{\mathrm{member}}$). Data points without error bars indicate cases where only a single group exists for a given $N_{\mathrm{member}}$, making it impossible to estimate a statistical uncertainty.}

    \label{fig3}
\end{figure}


The substantial increase in the number of Virgo Cluster members when adopting $\Delta V = 1000$\,km\,s$^{-1}$, despite a relatively small decrease in the total number of galaxy groups, arises from the reassignment of galaxies originally belonging to other rich systems. In particular, galaxies that were members of separate groups under the $\Delta V = 500$\,km\,s$^{-1}$ criterion are incorporated into the Virgo Cluster at the higher velocity threshold. 
For example, the second largest group identified with $\Delta V = 500$\,km\,s$^{-1}$—associated with the Ursa Major Cluster (UMa) - disappears from the group catalog when $\Delta V = 1000$,km,s$^{-1}$ is used, with its members subsumed in the Virgo Cluster. In contrast, the number of galaxy pairs (groups with two members) changes only slightly, from 267 to 258.

The galaxy groups identified in this study include most of the well-known groups and clusters in the local universe. The largest system is the Virgo Cluster, followed by the UMa Cluster, which contains 370 member galaxies. While the most common groups consist of only two galaxies, the richest group (excluding Virgo) contains up to 134 members. Of particular interest is a newly identified group centered on NGC~4168, located behind the Virgo Cluster. Although NGC~4168 is listed in the Extended Virgo Cluster Catalog \citep{kim14}, its distance suggests that it belongs to a distinct background system. Another notable system is the NGC~5846 group, situated in a relatively underdense region. Its association with the Virgo Cluster remains uncertain. It is unclear whether this group represents an independent system or a possible subgroup, like the NGC~5746 group, which was not identified in the present study.

\begin{figure}
	\centering
	\includegraphics[width=0.45\textwidth]{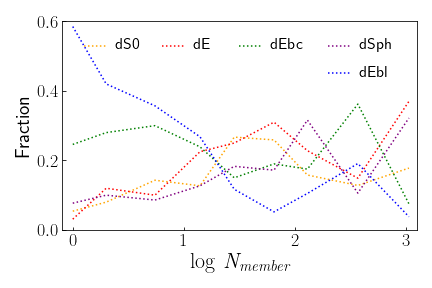}
	\caption{Fraction of dwarf galaxy types as a function of the number of group members. The dwarf types are adopted from CVCG \citep{ann15}: dS0 (dwarf lenticular), dE (dwarf elliptical), dE$_{\mathbf{bc}}$(blue-cored dwarf elliptical), dSph (dwarf spheroidal) and dE$_{\mathbf{b}}$(blue dwarf elliptical).}
	
	\label{fig4}
\end{figure}

The background density used in this study was calculated using the nearest-neighbour method with $n = 5$. This method depends on two parameters: the linking velocity ($\Delta V$) and the limiting absolute magnitude ($M_{\mathrm{lim}}$) for selecting neighbouring galaxies. We adopted $\Delta V = 500$\,km\,s$^{-1}$, consistent with the value used in our group-finding procedure. For the limiting magnitude, we used $M_{\mathrm{lim}} = -15.2$, which defines a volume-limited sample within $z = 0.01$. The computed background density was normalized by the mean surface density of the local universe, $\bar{\Sigma}$. Figure~\ref{fig3} displays the distribution of normalized background densities as a function of group richness (i.e., the number of member galaxies). The field environment corresponds to groups with $N = 1$. A detailed analysis of the physical properties of groups and clusters, particularly rich systems, will be presented in forthcoming papers.

Figure~\ref{fig4} presents the morphological fractions of the five early-type dwarf subtypes as a function of the group richness, measured by the number of member galaxies. In the field environment ($N = 1$), the population is dominated by blue dwarf ellipticals (dE$_{\mathbf{bl}}$). In contrast, the Virgo Cluster ($N_{\mathrm{member}} = 1068$) is overwhelmingly dominated by classical dwarf ellipticals (dE) and dwarf spheroidals (dSph). When considering only galaxy groups—excluding the Virgo and Ursa Major (UMa) Clusters—a clear trend emerges: the fractions of dS0, dE and dSph galaxies increase with group richness ($N_{\mathrm{member}}$).

It is noteworthy that the UMa Cluster ($N_{\mathrm{member}} = 370$) exhibits a markedly different morphological composition from the Virgo Cluster. In UMa, blue-cored dwarf ellipticals (dE$_{\mathbf{bc}}$) constitute the dominant population and the fraction of dE$_{\mathbf{bl}}$ galaxies slightly exceeds those of dS0, dE and dSph. In contrast, the Virgo Cluster contains significantly fewer dE$_{\mathbf{bc}}$ and dE$_{\mathbf{bl}}$ galaxies, with dE and dSph types being predominant.

\begin{figure*}
	\centering
	\includegraphics[width=0.8\textwidth]{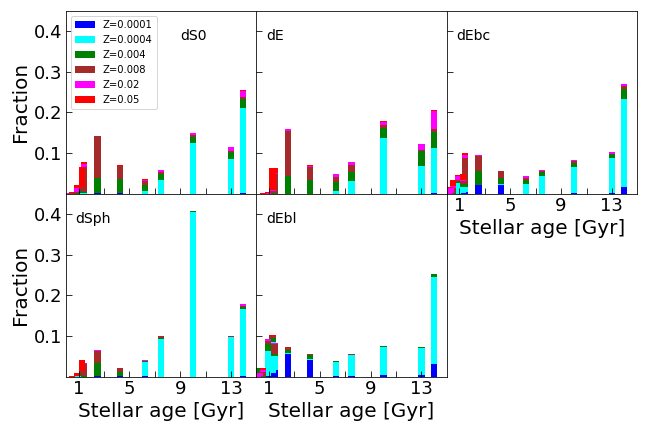}
	\caption{Stellar mass fractions as a function of stellar age, separated by morphological type. The contributions from different metallicities are color-coded as follows: Z = 0.0001 (blue), Z = 0.0004 (cyan), Z = 0.004 (green), Z = 0.008 (brown), Z = 0.02 (magenta) and Z = 0.05 (red). }
		   
	\label{fig5}
\end{figure*}

\section{Star Formation History}

\begin{figure*}
	\centering
	\includegraphics[width=0.8\textwidth]{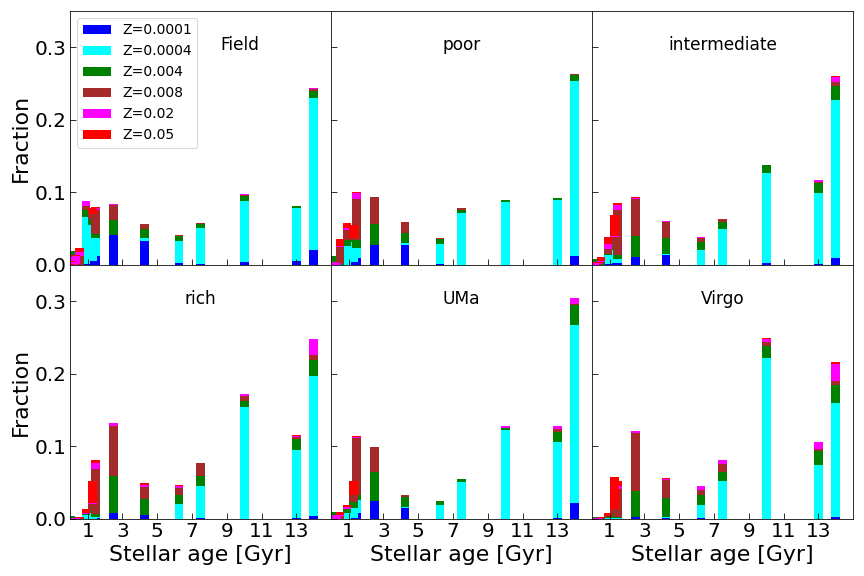}
	\caption{Star formation histories (SFHs) of early-type dwarf galaxies across six different environments: field, poor group, intermediate group, rich group, UMa Cluster, and Virgo Cluster. Each panel shows the distribution of stellar mass fraction as a function of stellar age (in Gyr). The vertical axis indicates the fraction of stellar mass formed at each age, normalised by the total stellar mass. The stacked bar charts are colour-coded by stellar metallicity, with metal-poor populations (low $Z$) at the bottom and metal-rich populations (high $Z$) at the top. This figure illustrates how the timing and chemical enrichment of star formation are shaped by environment.}
	\label{fig6}
\end{figure*}

The star formation histories of early-type dwarf galaxies can generally be divided into three distinct phases. The first phase is marked by an intense initial starburst followed by a secondary burst around 13~Gyr ago. The second phase, which began with a major episode of star formation around 10~Gyr ago, extends until approximately 5~Gyr ago. The third phase peaks around 2.5~Gyr ago and continues to the present day. The overall shape of these SFHs is influenced by both the morphology of the galaxy and the environment. As we demonstrate in the following subsections, the morphological type tends to exert a stronger influence than the environment. Figures~\ref{fig5} and Figure~\ref{fig6} summarize these aspects by presenting the metallicity-resolved stellar mass fractions of early-type dwarfs across different morphological subtypes and environmental categories.

\subsection{Morphology Dependence}

The SFHs of early-type dwarf galaxies exhibit two features common to all morphological types (see Figure~\ref{fig5}): (1) an early burst of star formation and (2) diminished bursts around 2.5~Gyr ago, which mark the peak of the third phase. In galaxies showing a burst of star formation at the beginning of the second phase, namely the dS0, dE, and dSph types, more than $\sim$ 70\% of the stellar mass is assembled during the first two phases, with negligible star formation activity after 1~Gyr ago. In contrast, dE$_{\mathbf{bc}}$ and dE$_{\mathbf{bl}}$ galaxies display pronounced star formation during the third phase, contributing up to $\sim$40\% of their total stellar mass, along with a substantial population of stars formed more recently than 1~Gyr ago. Their chemical enrichment histories also differ markedly from those of dS0, dE, and dSph galaxies. In the latter types, rapid chemical evolution during the third phase has led to the formation of predominantly metal-rich stars.

By comparison, a substantial fraction of stars formed during the third phase in dE$_{\mathbf{bc}}$ and dE$_{\mathbf{bl}}$ galaxies are metal-poor or even extremely metal-poor (Z = 0.0001). The differences between the two subtypes are also evident. In dE$_{\mathbf{bc}}$ galaxies, stars formed before the peak of the third phase exhibit a relatively even distribution of metallicities (Z = 0.0001, 0.004, and 0.008), whereas in dE$_{\mathbf{bl}}$ galaxies, stellar populations are dominated by extremely metal-poor stars (Z = 0.0001). Near the peak of the third star formation episode, dE$_{\mathbf{bc}}$ galaxies are characterized by a predominance of intermediate-metallicity stars (Z = 0.008), while dE$_{\mathbf{bl}}$ galaxies are dominated by stars with slightly lower metallicities (Z = 0.004). However, the metallicities of stars formed within the past $\sim$1~Gyr are similar in both subtypes.

Two key features in the star formation histories of early-type dwarf galaxies are particularly noteworthy: (1) the timing of peak starburst activity and (2) the relative paucity of extremely metal-poor stellar populations. In most early-type dwarfs, the peak of star formation occurred during the initial episode. However, dSph galaxies exhibit a distinct behaviour, with their maximum star formation activity occurring around 10~Gyr ago, rather than at the very beginning. This delayed onset of intense star formation in dSphs is likely attributable to their lower stellar masses compared to other subtypes. The vigorous starburst around 10~Gyr ago may have driven out a substantial fraction of their remaining gas, either into their haloes or expelled entirely from the systems. Consequently, the third phase of star formation in dSphs appears to have been significantly suppressed relative to that in dS0 and dE galaxies.

Extremely metal-poor stars are virtually absent in dS0, dE, and dSph galaxies, whereas they are present in non-negligible amounts in dE$_{\mathbf{bc}}$ and are abundant in dE$_{\mathbf{bl}}$ galaxies. In particular, the presence of such stars that formed during the initial starburst phase in dE$_{\mathbf{bc}}$ and dE$_{\mathbf{bl}}$ suggests that these galaxies originated in environments distinct from those of dS0, dE, and dSph types. The near absence of extremely metal-poor stars in the latter subtypes is generally attributed to pre-enrichment of the gas from which they formed. In contrast, the gas that gave rise to the galaxies dE$_{\mathbf{bc}}$ and dE$_{\mathbf{bl}}$ galaxies must have been significantly less enriched, implying a lower degree of chemical evolution in their formation environments. This interpretation is further supported by the continued formation of extremely metal-poor stars during the third star formation episode in these galaxies. Such stars likely originated from metal-free or nearly pristine cold gas accreted from the surrounding medium during later evolutionary stages.

In summary, the star formation history of dS0 galaxies closely resembles that of dE galaxies. In contrast, dE$_{\mathbf{bc}}$ galaxies are distinct not only from dS0 and dSph types but also from typical dE galaxies. The primary distinguishing feature of dE$_{\mathbf{bc}}$ galaxies is the presence of a substantial population of stars younger than 1~Gyr, indicating that star formation persisted almost to the present day. This extended star formation activity is also evident in dE$_{\mathbf{bl}}$ galaxies. The main difference between dE$_{\mathbf{bc}}$ and dE$_{\mathbf{bl}}$ galaxies lies in the metallicity of their intermediate-age stellar populations. In dE$_{\mathbf{bc}}$ galaxies, stars formed around 1~Gyr ago tend to have relatively high metallicities, whereas in dE$_{\mathbf{bl}}$ galaxies, stars of similar age are predominantly of lower metallicity (Z = 0.0004).

Despite these differences, the SFHs of dE$_{\mathbf{bc}}$ and dE$_{\mathbf{bl}}$ galaxies share some features with those of dS0, dE, and dSph galaxies. In all cases, metal-poor stars formed predominantly during the initial starburst phase. However, dE$_{\mathbf{bc}}$ and dE$_{\mathbf{bl}}$ galaxies differ in the metallicity of the stars formed during the third phase of star formation, prior to 1~Gyr ago. In dE$_{\mathbf{bc}}$ galaxies, comparable star formation rates are observed at 2.5 and 1.5~Gyr ago, whereas dE$_{\mathbf{bl}}$ galaxies exhibit a more pronounced peak at 1.5~Gyr ago.

\subsection{Environment Dependence}

We divided the galaxies in our sample into six environmental categories: field, poor, intermediate, and rich groups, defined by the number of group members; and the Ursa Major (UMa) and Virgo clusters. Poor groups contain only two members, intermediate groups up to seventy, and rich groups more than seventy. Although UMa is sometimes considered a massive group, we treat it as a cluster due to its richness and dynamical distinctiveness.

In the field environment (top-left panel of Figure~\ref{fig6}), early-type dwarf galaxies exhibit extended and multi-phased SFHs, shaped by intermittent episodes of star formation and chemical enrichment. The SFH begins with a strong initial burst within the first $\sim$1~Gyr, which contributes more than 35~per~cent of the total stellar mass. This phase is dominated by low-metallicity stars ($Z = 0.0004$), accompanied by smaller fractions of extremely metal-poor ($Z = 0.0001$) and mildly enriched ($Z = 0.004$) populations.

Following a brief quiescent period, a second episode of star formation occurred between $\sim$11 and 5~Gyr ago, contributing an additional $\sim$25~per~cent of the stellar mass. The metallicity distribution during this epoch remained similar to that of the initial burst, suggesting either slow chemical evolution or dilution by the inflow of pristine gas. In particular, an enhanced fraction of extremely metal-poor stars ($Z = 0.0001$) appears again around $\sim$4~Gyr ago, likely resulting from the accretion of metal-free gas from the intergalactic medium, a phenomenon rarely observed in denser environments.

The most recent episode, extending from $\sim$1~Gyr ago to the present, is characterized by chemically diverse star formation, notably including high-metallicity stars ($Z = 0.02$ and $Z = 0.05$). This trend suggests sustained chemical enrichment driven by both internal recycling and continued gas accretion, in the absence of strong environmental quenching.

In contrast, SFHs in group environments exhibit a systematic shift towards earlier and more metal-enriched star formation. As group richness increases, the contribution from recent star formation declines, and the prominence of extremely metal-poor stars around 3~Gyr ago diminishes, indicating a reduced inflow of pristine gas. Moreover, the stellar mass formed around 10~Gyr ago increases from poor to rich groups, suggesting that denser environments favour more efficient early star formation at the onset of the second phase.

The two cluster environments, UMa and Virgo, exhibit star formation histories broadly similar to that of the rich group environment. However, notable differences exist between the two clusters. The UMa Cluster, being dynamically young and relatively low in density, displays a strong initial burst of star formation, whereas the Virgo Cluster exhibits its highest peak at 10~Gyr ago. The presence of extremely metal-poor stars ($Z = 0.0001$) during the initial burst, along with a peak at 1.5~Gyr ago dominated by extremely metal-rich stars ($Z = 0.05$), makes the SFH of UMa resemble that of poor groups in both timing and chemical composition. It also retains evidence of extremely metal-poor star formation around 3~Gyr ago and shows modest recent activity, consistent with weak environmental quenching.

In contrast, Virgo displays its highest peak at 10~Gyr ago, along with a secondary peak at 2.5~Gyr ago, composed primarily of stars with intermediate metallicities ($Z = 0.004$ and $Z = 0.008$). Star formation at 1.5~Gyr ago is much reduced, yet the few stars formed during this epoch are predominantly extremely metal-rich ($Z = 0.05$). Recent star formation is rare but highly metal-enriched. Overall, Virgo appears to represent an evolved extension of group-like environments, both in temporal and chemical terms. The comparison between UMa and Virgo underscores the importance not only of environment, but also of its physical properties—such as density, mass, and dynamical state—in shaping the star formation and chemical evolution of galaxies.

\begin{figure}
	\centering
	\includegraphics[width=0.45\textwidth]{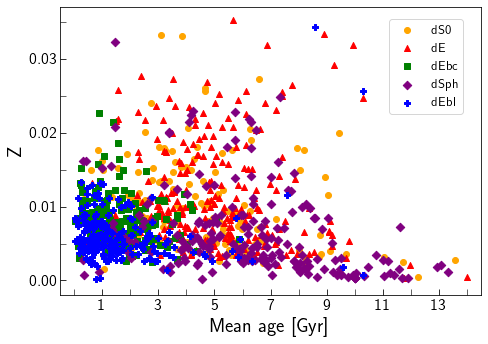}
	\caption{Luminosity-weighted mean metallicity ($Z$) as a function of luminosity-weighted mean stellar age. Morphological types are distinguished by different symbols and colors.}	
		
	\label{fig7}
\end{figure}

\subsection{Age and Metallicity}

One of the key advantages of reconstructing a galaxy’s star formation history through stellar population synthesis is that it enables the determination of the age and metallicity of the stellar populations contributing to the observed spectrum. Since the chemical evolution of a galaxy proceeds through successive generations of star formation and stellar evolution, the stellar populations within a galaxy typically exhibit a complex mixture of ages and metallicities. Population synthesis tools such as STARLIGHT analyze the composite stellar spectrum to infer the distribution of ages and metallicities of the contributing stars. From this analysis, the luminosity-weighted average age and metallicity of the stellar population can be derived. Figure~\ref{fig7} presents the relationship between the mean stellar age and average metallicity for early-type dwarf galaxies, classified by morphological type. These quantities were computed from the data shown in Figure~\ref{fig5}, using the equations provided below.

\begin{equation}
	\log t_{\ast} = \sum_{j=1}^{N_{\ast}} L_{j} \log t_{j}
\end{equation}

\begin{equation}
	Z = \sum_{j=1}^{N_{\ast}} L_{j} Z_{j}
\end{equation}

\noindent{where $L_{j}$ is the fractional luminosity contribution of the $j$-th stellar population and $\log t_{j}$ and $Z_{j}$ denote its stellar age and metallicity, respectively.}
Several trends are evident in Figure~\ref{fig7}. First, the mean stellar ages of dS0, dE and dSph galaxies generally exceed 3 Gyr, with dSph galaxies hosting the oldest stellar populations. In contrast, dE$_{\mathbf{bc}}$ and dE$_{\mathbf{bl}}$ galaxies exhibit significantly younger populations, with most having mean stellar ages below 3 Gyr and some even younger than 1 Gyr.

In terms of metallicity, dS0, dE and dSph galaxies span a broad range, from $Z = 0.000$ to $Z = 0.035$, though only a small fraction exhibit metallicities greater than $Z = 0.030$. On the other hand, the dE$_{\mathbf{bl}}$ galaxies predominantly have metallicities below $Z = 0.010$, while the dE$_{\mathbf{bc}}$ galaxies are largely concentrated around $Z = 0.010$. This suggests that, on average, both blue early-type dwarf populations possess sub-solar metallicities—approximately half that of the Sun.

The chemical evolution of galaxies generally proceeds toward higher metallicity as they age, since successive generations of star formation enrich the interstellar medium with heavy elements. As such, galaxies with younger mean stellar ages are expected to exhibit higher average metallicities. This trend is clearly observed in Figure~\ref{fig7} when the sample is divided by morphological type.

Although the slope of the metallicity–age relation varies among galaxy types, a consistent pattern emerges: metallicity increases with decreasing stellar age. The trend is most pronounced in the dS0 and dE galaxies, where the enrichment occurs relatively rapidly. In comparison, dSph galaxies show a more gradual increase in metallicity, suggesting a slower chemical evolution despite their older stellar populations.

In contrast, the dE$_{\mathbf{bc}}$ and dE$_{\mathbf{bl}}$ galaxies exhibit notably slower chemical evolution. This is particularly evident for dE$_{\mathbf{bl}}$ galaxies, whose metallicities remain largely below $Z = 0.010$, indicating a distinct evolutionary path from the other types. This slow enrichment is likely due not only to a more extended or inefficient star formation history, but also to the infall of metal-poor gas during a secondary star formation episode, as previously discussed. Stars formed from this externally accreted gas would naturally reflect lower metallicities, contributing to the observed trend.

\begin{figure*}
		\centering
		\includegraphics[width=0.83\textwidth]{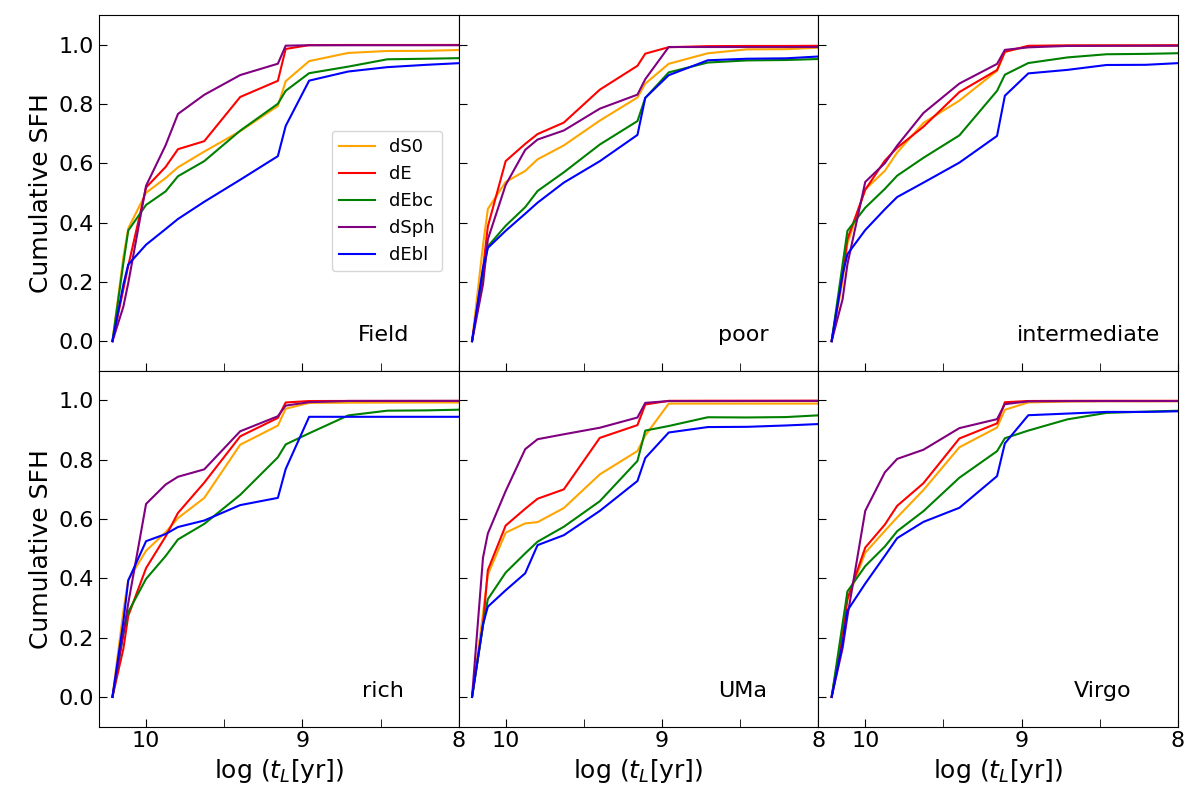}
		\caption{Cumulative star formation histories (cSFHs) of early-type dwarf galaxies in different group environments. Morphological types are distinguished by colors, as indicated in the legend.
		}
		\label{fig8}
\end{figure*}

\section{Cumulative Star Formation History}

The cumulative star formation history (cSFH) traces the integrated fraction of stellar mass formed over cosmic time, in contrast to the SFH, which quantifies the rate of star formation as a function of time. By construction, cSFHs provide a smoothed and monotonic view of stellar mass assembly, making them particularly advantageous for comparative analyses across galaxy types and environments. Whereas SFHs can be subject to short-term stochastic fluctuations, cSFHs offer a more robust framework for investigating long-term evolutionary trends, especially in low-mass systems such as early-type dwarf galaxies. There are two characteristic times $\tau_{50}$ and $\tau_{90}$. They denote the lookback time when 50\% and 90\% of the stellar mass had formed, respectively. The former is considered formation time, and the latter is used for the quenching time \citep{wei15}. \citet{rom24a} devised two time scales $\Delta\tau_{50}$ and $\Delta\tau_{90}$ using $\tau_{50}$ and $\tau_{90}$ together with the lookback time to the Big Bang, $\tau_{BB}$. We first describe the cSFHs of early-type dwarfs and then analyze the characteristic times derived from the cSFHs.


\subsection{Morphology Dependence at Fixed Environment}

Figure~\ref{fig8} illustrates how the cSFH of early-type dwarf galaxies evolves as a function of log~$t_{L}$ where $t_L$ is the lookback time expressed in years, across different environments. The cSFHs of different subtypes look quite different, except for the following cases where little difference is seen: dS0 and dE$_{\mathbf{bc}}$ in the field environment; dE$_{\mathbf{bc}}$ and dE$_{\mathbf{bl}}$ in the poor group; dS0, dE, and dSph in the intermediate group; dS0 and dE in the rich group; dE$_{\mathbf{bc}}$ and dE$_{\mathbf{bl}}$ in the UMa Cluster; and dS0 and dE in the Virgo Cluster. The apparently large differences between the cSFHs of other subtypes are statistically significant according to the Kolmogorov-Smirnov test.

In the field environment, dSph galaxies exhibit the most rapid early star formation, followed by dE galaxies. In contrast, dE$_{\mathbf{bl}}$ galaxies show the slowest star formation history, having formed only about 30\% of their stellar mass by log~$t_L=10$ (i.e., 10 Gyr ago), with star formation continuing until recent times.

Interestingly, the cSFHs of dS0 and dE$_{\mathbf{bc}}$ galaxies, which possess central lens-like structures and prominent blue cores, respectively, are nearly identical in the field environment. While both types experienced rapid initial star formation, they were soon overtaken in cumulative mass growth by dSph and dE galaxies. A marked change in the star formation rate is observed around log~$t_L\approx9$ for most galaxy types, with the transition being particularly pronounced in dE$_{\mathbf{bl}}$ galaxies. Notably, dSph galaxies had already formed approximately 80\% of their present-day stellar mass by log~$t_L\approx9.8$, highlighting their extremely early and rapid star formation. These galaxies are likely remnants of primordial systems that formed in the early universe.

In poor group environments, the relative differences in star formation rates among early-type dwarf subtypes resemble those observed in the field. However, several notable deviations from the field population are apparent. Most significantly, the star formation history of dS0 galaxies differs: While star formation proceeds relatively rapidly, it extends until log~$t_L\approx8.5$, which is later than that of the dSph and dE galaxies in the same environment. For dE$_{\mathbf{bc}}$ galaxies, the initial burst of star formation reaches only about 80\% of that seen in the field, resulting in a slope similar to the field counterpart during early epochs. Nonetheless, their overall star formation progresses more slowly, eventually converging with that of dE$_{\mathbf{bl}}$ galaxies around log~$t_L\approx9$. Among the five morphological types, dE$_{\mathbf{bl}}$ galaxies in poor groups exhibit the star formation history most closely aligned with their counterparts in the field, suggesting that their evolutionary pathways are less sensitive to environmental differences compared to the other types.

	\begin{figure*}
		\centering
		\includegraphics[width=0.83\textwidth]{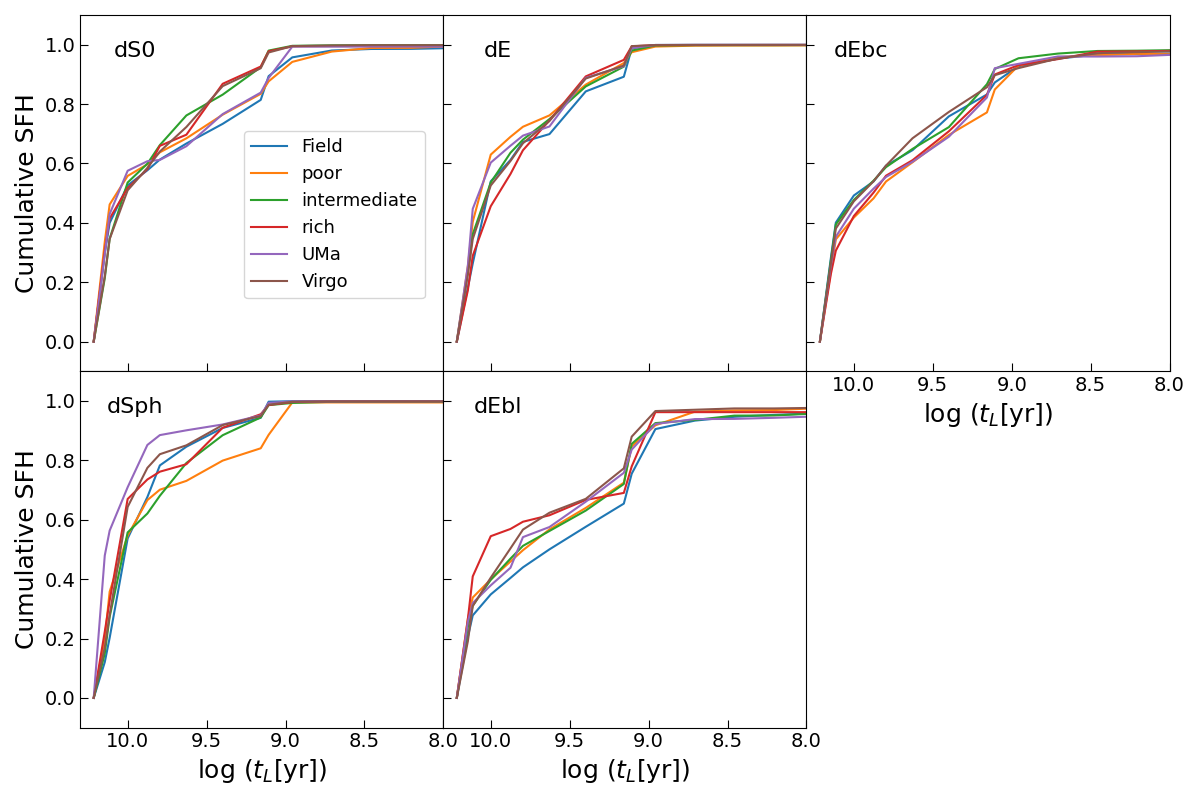}
		\caption{Cumulative star formation histories (cSFHs) of early-type dwarf galaxies, divided by morphological type. Different environments are distinguished by colors, as indicated in the legend.}
		
		\label{fig9}
	\end{figure*}

The top-right panel of Figure~\ref{fig8} presents the cSFHs of galaxies in intermediate group environments, which show marked differences compared to those in the field and poor groups. The most striking feature is the convergence of star formation histories among dS0, dE and dSph galaxies: with the exception of dE$_{\mathbf{bc}}$ and dE$_{\mathbf{bl}}$, all types exhibit nearly identical cSFHs, suggesting strong environmental regulation. The cSFH of dE$_{\mathbf{bc}}$ galaxies in intermediate groups closely resembles that of their counterparts in the field and in poor groups. Similarly, dE$_{\mathbf{bl}}$ galaxies show a comparable overall pattern but with somewhat more rapid early star formation compared to those in the field. In rich group environments (bottom-left panel), dSph galaxies initially form more stars than dE galaxies, a reversal of the trend seen in less dense environments. Moreover, dE$_{\mathbf{bl}}$ galaxies in rich groups exhibit relatively rapid early star formation, forming approximately 55\% of their stellar mass by log~$t_L = 10$. After this early phase, their star formation slows considerably until log~$t_L \approx 9$, at which point a distinct secondary burst of star formation occurs. This late-time enhancement in star formation at log~$t_L \approx 9$ is also evident in the field, though it appears less pronounced in the poor and intermediate groups. The continuation of star formation beyond log~$t_L \approx 9$ in dE$_{\mathbf{bc}}$ and dE$_{\mathbf{bl}}$ galaxies is consistent across environments, suggesting that their late evolutionary phases are less sensitive to the environment than their early star formation histories.

The bottom-middle panel of Figure~\ref{fig8} presents the cSFHs for the Ursa Major (UMa) Cluster, which contains approximately 370 member galaxies, making it the second largest group/cluster in the sample after the Virgo Cluster. Although the overall trends in cSFH in UMa resemble those seen in rich group environments, several notable differences are observed. First, dSph galaxies in UMa exhibit even more rapid early star formation compared to those in rich groups, resulting in a greater contrast with the cSFH of dE galaxies. In contrast, dE$_{\mathbf{bl}}$ galaxies in UMa show relatively limited early star formation, with cumulative mass increasing steadily until log~$t_L = 9$. The dE$_{\mathbf{bc}}$ galaxies form stars slightly more rapidly than dE$_{\mathbf{bl}}$, but their overall star formation histories are broadly similar. The most distinctive feature in the UMa Cluster is the star formation history of dS0 galaxies. Unlike their counterparts in intermediate and rich groups—where dS0 galaxies closely track the cSFHs of dE galaxies—those in UMa show a markedly slower evolution after log~$t_L = 10$. The slope of the cSFH flattens noticeably, indicating a significant decline in the star formation rate during later epochs.

The final panel (bottom-right) of Figure~\ref{fig8} shows the cSFHs of early-type dwarf galaxies in the Virgo Cluster—the largest galaxy cluster in the local universe. The overall star formation histories of Virgo Cluster galaxies are broadly similar to those observed in the UMa Cluster. However, notable distinctions remain. For instance, the cSFH of dS0 galaxies in Virgo more closely resembles that of dS0 galaxies in rich groups than those in UMa. Their cumulative stellar mass growth closely tracks that of dE galaxies and less than 50\% of their present-day stellar mass was assembled by log~$t_L = 10$. This suggests that most dE and dS0 galaxies in the Virgo Cluster are unlikely to be primordial in origin; instead, they were likely transformed from late-type progenitors through environmental mechanisms.
In contrast, dSph galaxies in Virgo formed approximately 80\% of their stellar mass by log~$t_L = 10$, indicating that many of them are plausible relics of primordial systems. Another noteworthy feature is the cSFH of dE$_{\mathbf{bl}}$ galaxies, which more closely resembles that of galaxies in rich groups than in the UMa Cluster. In particular, star formation in dE$_{\mathbf{bl}}$ galaxies appears to have been significantly suppressed during the second phase of their evolutionary history, implying a stronger environmental influence on their late-time evolution.

	\begin{figure*}
		\centering
		\includegraphics[width=0.83\textwidth]{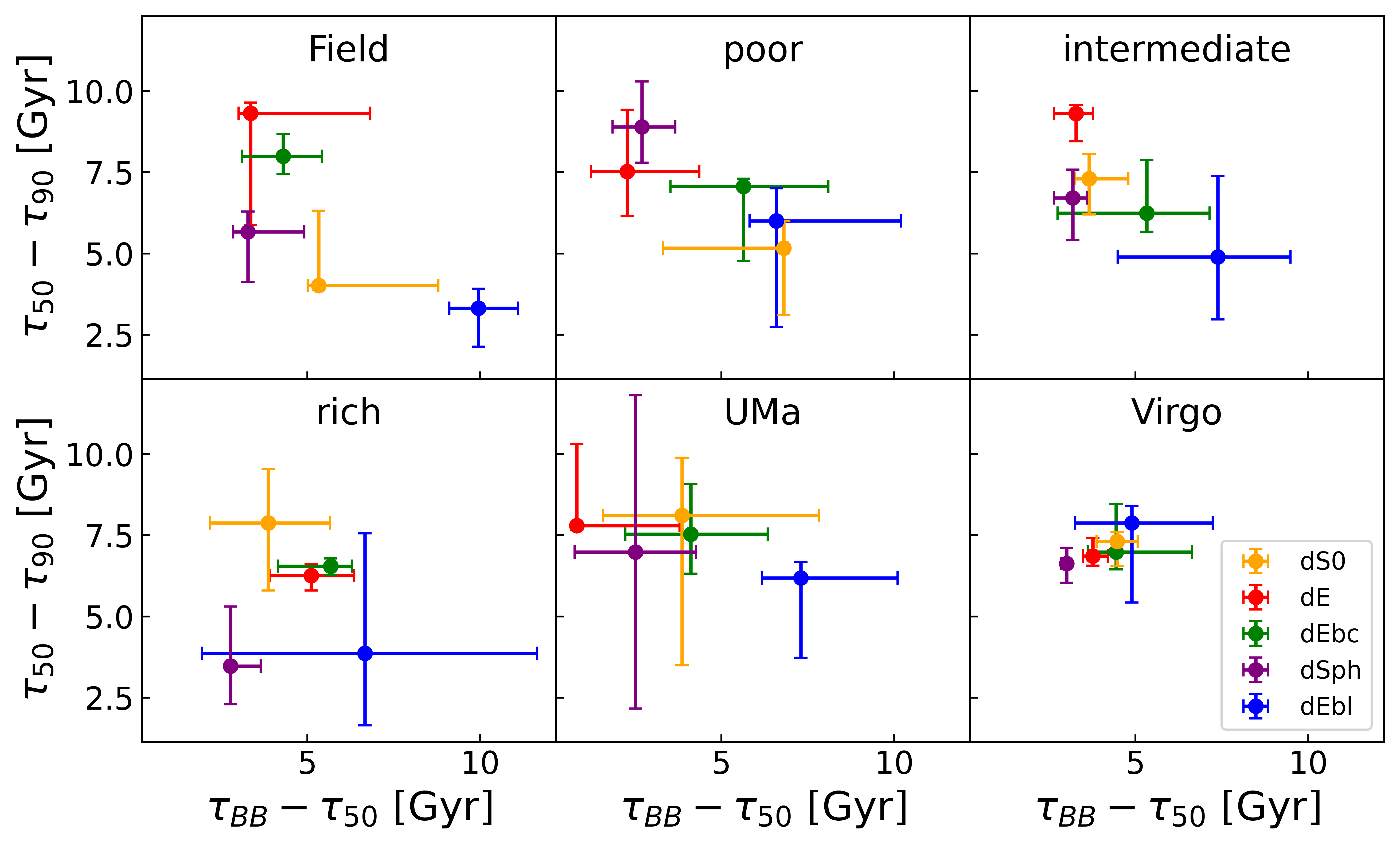}
		\caption{Star formation histories of early-type dwarf galaxies by environment. Each panel shows the median values of $\tau_{\mathrm{BB}} - \tau_{50}$ (x-axis) and $\tau_{50} - \tau_{90}$ (y-axis) for different morphological subtypes (coloured symbols). Here, the terms $\tau_{50}$ and $\tau_{90}$ indicate the lookback times when 50\% and 90\% of the stellar mass had formed, respectively, while $\tau_{BB}$ denotes the lookback time to the Big Bang. Error bars denote 68\% bootstrap confidence intervals.
		}
		
		\label{fig10}
	\end{figure*}
	%

	\begin{figure*}
		\centering
		\includegraphics[width=0.83\textwidth]{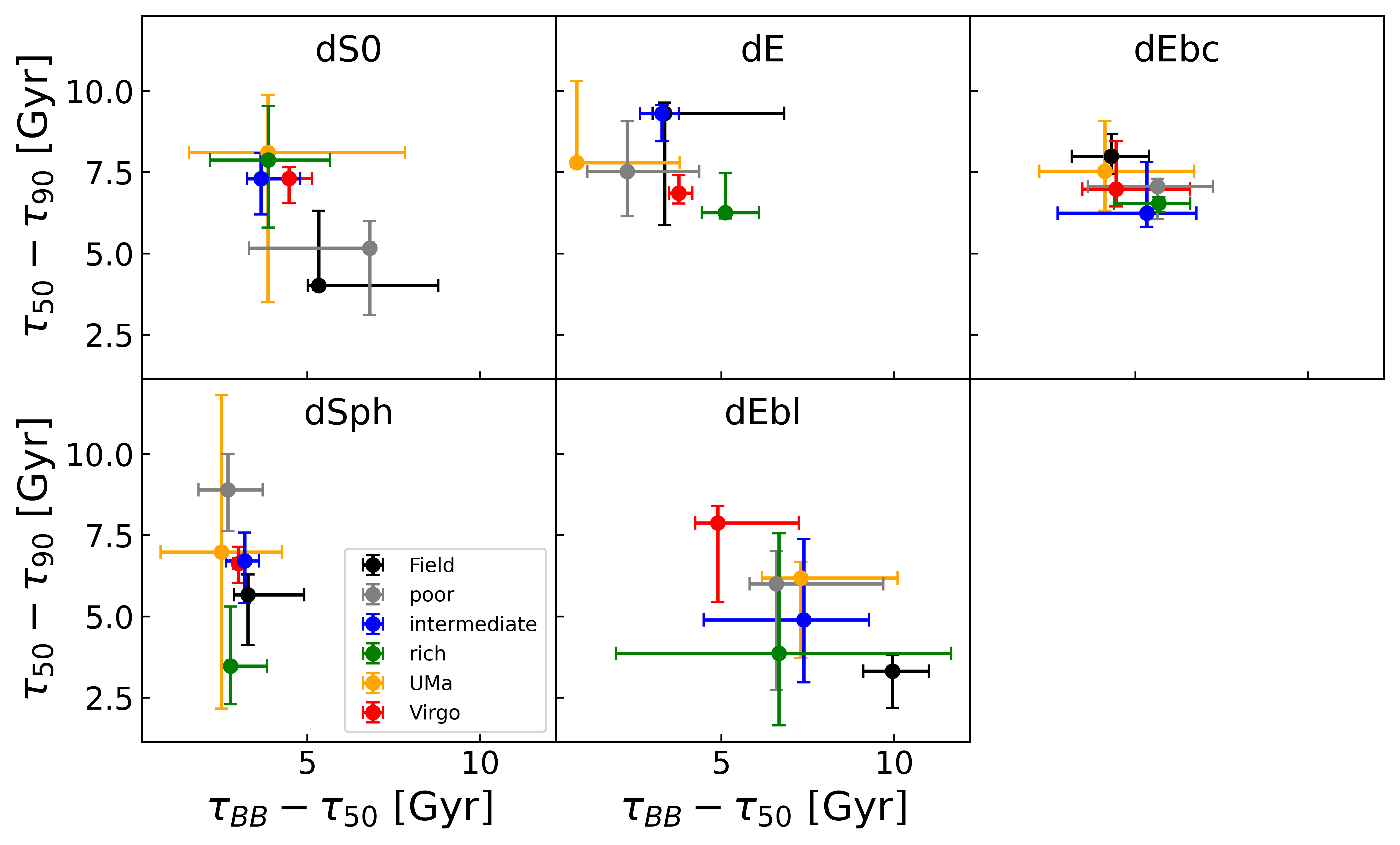}
		\caption{Star formation histories of early-type dwarf galaxies by morphological subtype. Each panel shows the median values of $\tau_{\mathrm{BB}} - \tau_{50}$ (x-axis) and $\tau_{50} - \tau_{90}$ (y-axis) for a given subtype, with different environments distinguished by colour. Error bars indicate the 68\% bootstrap confidence intervals.
		}
		
		\label{fig11}
	\end{figure*}

\subsection{Environmental Dependence at Fixed Morphology}

Figure~\ref{fig9} displays the cSFHs for each morphological subtype, now separated by environment. This allows for a direct assessment of how the environment influences the stellar mass assembly of galaxies with similar intrinsic structures. The impact of environment is heterogeneous across morphological types. Whereas cSFHs differ considerably between subtypes in a fixed environment, they tend to remain relatively similar for a given subtype in different environments. However, dS0 and dSph show pronounced environmental differences that are statistically significant, as confirmed by the Kolmogorov-Smirnov test. For example, dS0 galaxies in the poor group show statistically significant differences compared to those in the intermediate group and Virgo Cluster, with $p$-values of 0.0019 and 0.0056, respectively. The dE galaxies exhibit a significant difference between the poor group and the UMa Cluster, with $p = 0.035$. In particular, the cSFH of dSph galaxies in the poor group differs significantly not only from that in the low-density field environment, but also from those in rich groups and clusters such as UMa and Virgo, with $p < 0.014$.

For dS0 galaxies, approximately 60\% of their stellar mass is formed by log~$t_L = 10$ in all environments. Star formation typically ends around log~$t_L \approx 9$, with the exception of galaxies in the field and poor groups, where star formation persists slightly longer. 
dE galaxies generally exhibit more intense early star formation than dS0 galaxies, with a greater fraction of stellar mass formed by log~$t_L = 10$, except in rich groups. In all environments, star formation in dE galaxies concludes around log~$t_L \approx 9$.

The dE$_{\mathbf{bc}}$ galaxies, in contrast, show the least active early star formation among the five types. By log~$t_L = 10$, only about 50\% of their current stellar mass has formed. While their star formation typically ends before log~$t_L \approx 9$, as with dE galaxies, some dE$_{\mathbf{bc}}$ galaxies continue forming stars until recent times. This ongoing activity stands in contrast to dE and dSph galaxies, where virtually no star formation occurs after log~$t_L \approx 9$.

As previously hinted in Figure~\ref{fig8}, dSph galaxies exhibit predominantly early star formation in rich groups, UMa and Virgo where more than ~60\% of their stellar mass is formed by log~$t_L = 10$, indicating that the majority of dSph systems in these environment are likely primordial in origin. This trend is especially pronounced in the UMa Cluster, where nearly 90\% of star formation occurs during the earliest phase of evolution. This is likely due to the rapid consumption and subsequent removal of gas—possibly driven by environmental processes—following the initial starburst, which suppressed any further star formation.

The cSFHs of dE$_{\mathbf{bl}}$ galaxies also exhibit significant dependence on environment. In rich group environments, they show the most active early star formation among all environments, forming nearly 50\% of their present-day stellar mass by log~$t_L = 10$. However, star formation declines sharply between log~$t_L \approx 10$ and log~$t_L \approx 9$, followed by a sudden and pronounced resurgence at log~$t_L \approx 9$. This secondary burst of star formation is unique to dE$_{\mathbf{bl}}$ galaxies and is observed across multiple environments, although the strength of the burst varies. This unusual late-time star formation activity may be driven by the accretion of external gas. As shown in Figure~\ref{fig5}, many of the stars formed during this second phase originated from gas with low metallicity, supporting the idea that this gas was acquired externally.

Another notable feature of dE$_{\mathbf{bl}}$ galaxies is the persistence of low-level star formation over the past billion years, albeit with environmental variation. This ongoing star formation likely accounts for their overall blue appearance. Although stars formed recently contribute only a small fraction to the total stellar mass, their high luminosities—especially from massive young stars—dominate the integrated light, resulting in a blue color. It is also worth noting that dE$_{\bold{bl}}$ galaxies in the field exhibit the slowest star formation rates during the early phase (from log~$t_L \approx 10$ to log~$t_L \approx 9$), further highlighting the environmental sensitivity of their evolutionary histories.

\subsection{Galaxy Formation Time Scales of Early-Type Dwarf Galaxies}

Considering both morphology and environment, the above results support a dual-dependence scenario in which star formation histories of early-type dwarf galaxies are shaped by both factors. The morphological subtype reflects intrinsic formation timescales and mechanisms, whereas the environment modulates these evolutionary pathways to varying degrees. The cSFH framework thus proves especially effective in revealing such nuanced dependencies.

\subsubsection{Role of Morphology}

Fig.~\ref{fig10} presents median values and 68\% confidence intervals for two characteristic star formation timescales of early-type dwarf galaxies across six environments, grouped by morphological subtype. The medians were computed using iterative 3$\sigma$ clipping, and the confidence intervals were obtained from 1000 bootstrap resamplings. The figure highlights the interplay between morphology and environment in shaping the SFHs of early-type dwarfs, with each panel corresponding to a fixed environment, from the field to the Virgo Cluster, and color-coded symbols denoting morphological subtypes. Several consistent patterns emerge:

dE galaxies tend to exhibit short $\tau_{\mathrm{BB}} - \tau_{50}$ and long $\tau_{50} - \tau_{90}$ timescales, particularly in low-density environments such as the field and intermediate groups. This indicates a rapid onset of star formation followed by an extended and gradual decline, suggesting a relatively early formation with prolonged star-forming activity.

dE$_{\mathrm{bl}}$ galaxies, in contrast, consistently display long $\tau_{\mathrm{BB}} - \tau_{50}$ and short $\tau_{50} - \tau_{90}$ values, implying a delayed but rapid assembly of stellar mass. This trend is especially prominent in the field and poor groups, where gas accretion and delayed collapse may have facilitated late, concentrated star formation.

dSph galaxies show large scatter, but generally feature short $\tau_{\mathrm{BB}} - \tau_{50}$ and moderate $\tau_{50} - \tau_{90}$. This suggests an early onset with relatively rapid cessation, possibly driven by environmental quenching or low gas retention efficiency.

dS0 galaxies often exhibit intermediate to long $\tau_{\mathrm{BB}} - \tau_{50}$ and broad $\tau_{50} - \tau_{90}$ distributions. This diversity implies multiple evolutionary pathways, with lens-like morphology possibly resulting from either extended star formation or morphological transformation.

dE$_{\mathrm{bc}}$ galaxies tend to have short $\tau_{\mathrm{BB}} - \tau_{50}$ and relatively long $\tau_{50} - \tau_{90}$ in most environments, suggesting that while their initial formation was early, star formation persisted over extended periods—consistent with their blue cores indicating recent or residual activity.

Notably, the variation among subtypes is more distinct in low-density environments (e.g., field and poor groups), whereas in rich groups and clusters (e.g., Virgo), these differences become less pronounced. This convergence suggests that dense environments act to truncate or homogenize star formation histories across morphological types.

Overall, this diagnostic plot offers an intuitive yet quantitative framework for comparing SFHs across galaxy types and environments. It supports the view that both \textit{nurture} (environment) and \textit{nature} (morphology) contribute to the evolutionary history of early-type dwarfs, with environment playing an increasingly dominant role in denser regions of the Universe.

\subsubsection{Role of Environment}

Fig.~\ref{fig11} presents the environmental dependence of star formation timescales for early-type dwarf galaxies, shown separately for each morphological subtype. Each panel corresponds to a fixed subtype—dS0, dE, dE${\mathrm{bc}}$, dSph, and dE${\mathrm{bl}}$—and within each panel, coloured points represent the median values of $\tau_{\mathrm{BB}} - \tau_{50}$ and $\tau_{50} - \tau_{90}$ in different environments. The error bars denote the confidence intervals 68\% derived from 1000 bootstrap resamplings, and the medians were computed using iterative 3$\sigma$ trimming.

Among the subtypes, dS0 galaxies exhibit the most striking environmental variation. In low-density environments such as the field and poor groups, they tend to experience delayed but rapid star formation (long $\tau_{\mathrm{BB}} - \tau_{50}$ and short $\tau_{50} - \tau_{90}$), while in dense environments such as Virgo and UMa, their star formation starts earlier and proceeds over longer durations. This reversal of both the onset and the duration suggests that dS0s are highly sensitive to environmental conditions and may form through different channels depending on the local density.

dE galaxies also show diverse star formation histories across environments. In UMa, they form the earliest—with the shortest $\tau_{\mathrm{BB}} - \tau_{50}$—and complete most of their star formation over moderate durations. In contrast, dEs in the rich group and Virgo begin to form later, and those in the intermediate group and field undergo the most prolonged star formation episodes, as evidenced by their large $\tau_{50} - \tau_{90}$. These variations indicate that the onset and duration of star formation in dEs are strongly modulated by the environment, with dense regions that promote earlier formation.

dE$_{\mathrm{bc}}$ galaxies show remarkably uniform timescales across environments, implying that their prolonged star formation may be driven more by internal processes (e.g., residual gas retention or feedback regulation) than by external environmental effects.

dSph galaxies exhibit remarkably consistent star formation onset times across all environments, with $\tau_{\mathrm{BB}} - \tau_{50}$ values clustering between 2.5 and 3 Gyr. This suggests that their formation began early and was largely independent of environmental density. However, their star formation durations, as traced by $\tau_{50} - \tau_{90}$, vary significantly with environment. In poor groups, the duration extends up to $\sim$8.5 Gyr, indicating long-lasting star formation. By contrast, dSphs in rich groups show much shorter durations, just above 3 Gyr. Interestingly, intermediate groups yield durations comparable to those in UMa and Virgo ($\sim$7 Gyr), while field galaxies fall in between, with $\tau_{50} - \tau_{90}$ around 6 Gyr. These trends imply that, although dSphs formed early in all environments, their subsequent star formation histories were strongly modulated by environment, particularly in terms of quenching timescales.

dE$_{\mathrm{bl}}$ galaxies exhibit strong environmental variation in their star formation histories. In the field, they tend to show delayed but rapid formation, with long $\tau_{\mathrm{BB}} - \tau_{50}$ and short $\tau_{50} - \tau_{90}$ timescales, indicating a late onset followed by a compressed period of star formation. In contrast, their counterparts in high-density environments such as Virgo and UMa formed earlier and over more extended durations, suggesting prolonged star formation activity. In rich groups, dE$_{\mathrm{bl}}$ galaxies show intermediate behavior: they tend to form relatively early but complete most of their star formation over a shorter timescale. This diversity highlights the sensitivity of dE$_{\mathrm{bl}}$ galaxies to environmental effects such as ram-pressure stripping or tidal interactions, which may accelerate or suppress star formation depending on the local density.

In summary, taken together, these results demonstrate that the influence of environment on the star formation histories of early-type dwarf galaxies is highly dependent on the morphological subtype. dS0 and dE$_{\mathrm{bl}}$ galaxies show the strongest environmental modulation, with both the timing and duration of star formation varying systematically with local density. dE galaxies also respond strongly to environmental conditions, particularly in the timing of star formation onset. In contrast, dE$_{\mathrm{bc}}$ galaxies appear to be largely insensitive to the environment, suggesting that internal mechanisms dominate their extended star formation. dSph galaxies present a hybrid case: their formation onset is largely invariant across environments, but their star formation durations vary significantly, implying environmental regulation of quenching efficiency. These findings highlight that the environment plays a multifaceted role in shaping early-type dwarf evolution, either by altering the onset of star formation, regulating its duration, or suppressing it altogether, and that the relative importance of these processes differs by morphological subtype.

\begin{figure*}
	\centering

	\includegraphics[width=0.9\textwidth]{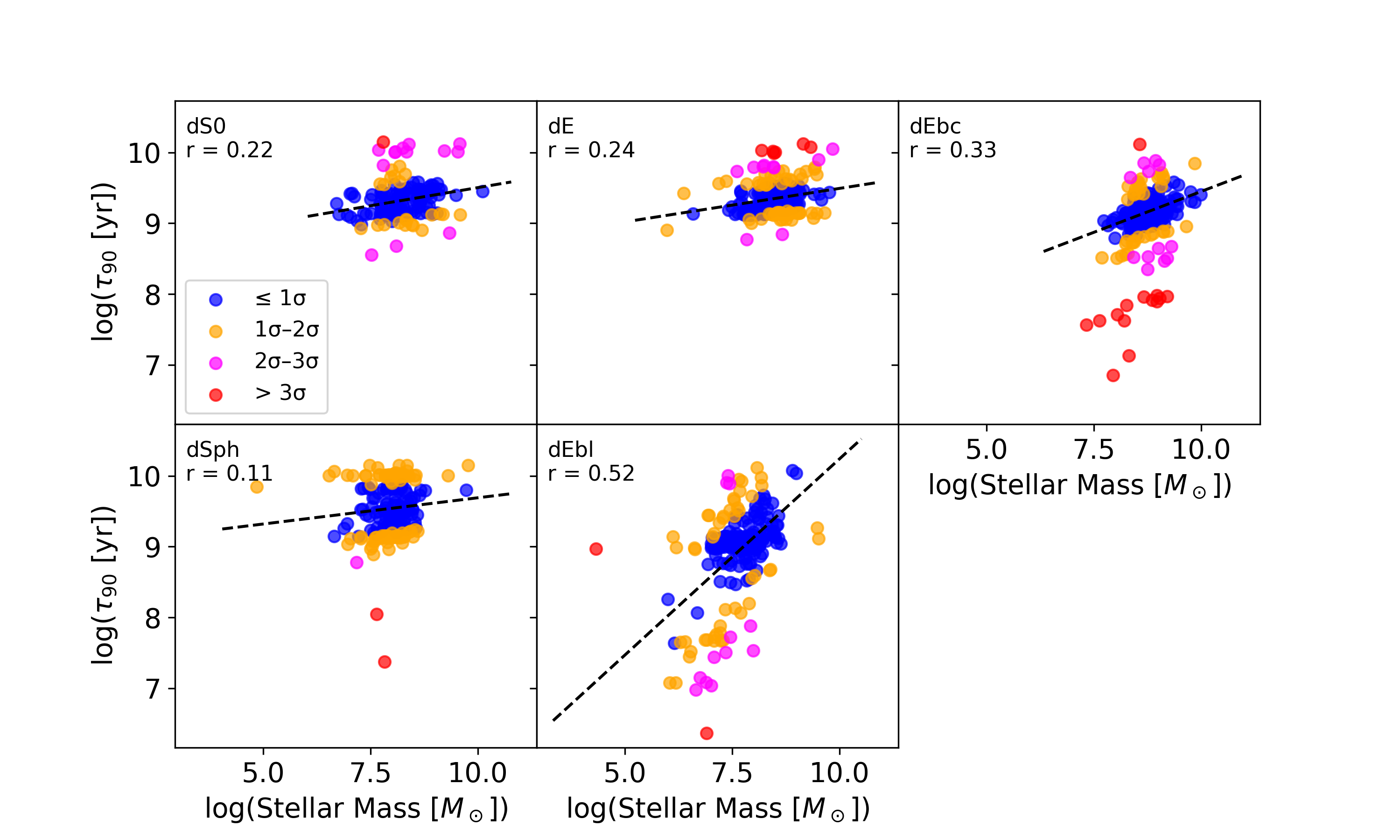}
	\caption{Quenching time ($\log(\tau_{90})$) versus stellar mass for each subtype of early-type dwarf galaxies. The black short-dashed lines indicate the regression fits with $3\sigma$ clipping for each subtype, while the green line shows the global regression fit. Pearson correlation coefficients ($r$) are shown below each subtype label. Data points are color-coded according to their deviation from the global regression line, as indicated in the legend in the lower-left panel.}
	
	\label{fig12}
\end{figure*}

\subsection{Quenching Epoch}

Figure~\ref{fig12} shows the relationship between $\tau_{90}$ and stellar mass for early-type dwarf galaxies, subdivided by morphological type. For each type, we show the individual regression line and the corresponding Pearson correlation coefficient ($r$). The global regression line derived from the combined galaxy sample is also shown for comparison. All regression lines were determined using 3$\sigma$ clipping to mitigate the influence of outliers. As evident in the figure, galaxies with higher stellar mass tend to cease star formation earlier. However, the weak slope of the global regression line (0.064 ± 0.033) indicates that the dependence of the quenching time on the mass is generally mild. Additionally, the correlation between $\tau_{90}$ and stellar mass is weak across most types, as reflected by the low r values, with the exception of dE$_{\mathbf{bl}}$ galaxies.

The slopes of the regression lines for three red early-type dwarf populations, dS0 ($0.103 \pm 0.074$), dE ($0.095 \pm 0.050$) and dSph ($0.074 \pm 0.097$), are similar to that of the global trend. In contrast, the two blue early-type dwarfs exhibit significantly steeper slopes: $0.231 \pm 0.101$ for dE$_{\mathbf{bc}}$ and $0.555 \pm 0.124$ for dE$_{\mathbf{bl}}$. The unusually steep slope of the dE$_{\mathbf{bl}}$ population arises from galaxies with extremely delayed quenching, some of which are still forming stars. These galaxies lie within 3$\sigma$ of the regression line and thus contribute to its steepness. For the dE$_{\mathbf{bc}}$ galaxies, however, many of the late-quenching members are excluded by the 3$\sigma$ clipping due to their large deviations, resulting in a slope less than half that of the dE$_{\mathbf{bl}}$ population.

Another notable feature in Figure~\ref{fig12} is the early quenching observed in dS0, dE and dSph galaxies, all of which exhibit red colors. Star formation in these systems largely ceased at log$\tau_{90}$ $\gtrsim9$, with mean values of log~$\tau_{90}$ measured as $9.33\pm0.02$ for dS0, $9.37\pm0.02$ for dE and $9.52\pm0.03$ for dSph galaxies. Welch’s t-test indicates that the mean quenching time of dSph galaxies is statistically distinct from those of dS0 and dE galaxies, whereas the difference between dS0 and dE is not statistically significant.

The early quenching of dSph galaxies suggests that a substantial fraction of them may be relics of primordial systems formed in the early universe. It is also plausible that some dS0 and dE galaxies originated from such early-forming populations, as several exhibit extremely early quenching with log$\tau_{90}$ $\gtrsim 10$, as shown in Figure~\ref{fig8}. However, a considerable number of dS0 and dE galaxies quenched more recently, with quenching times below their respective means. These galaxies are likely the result of morphological transformation from late-type progenitors. A likely mechanism for such a transformation is the removal of ram pressure \citep{gg72} and gravitational harassment \citep{moo96}, both of which are effective in dense environments. These processes can remove the interstellar medium and induce structural changes via tidal interactions, resulting in a cessation of star formation and a transition to spheroidal morphology.

In contrast, the two early-type dwarfs showing blue features, dE$_{\mathbf{bc}}$ and dE$_{\mathbf{bl}}$, exhibit relatively late quenching, with mean log$\tau_{90}$ values of $9.06\pm0.04$ and $8.92\pm0.04$, respectively. This delayed quenching is driven by a significant number of galaxies with ongoing star formation, leading to $\tau_{90}$ values as low as log$\tau_{90} \approx 8$.

\subsection{Virgo Cluster} 

\begin{figure*}
			\centering
			\includegraphics[width=0.8\textwidth]{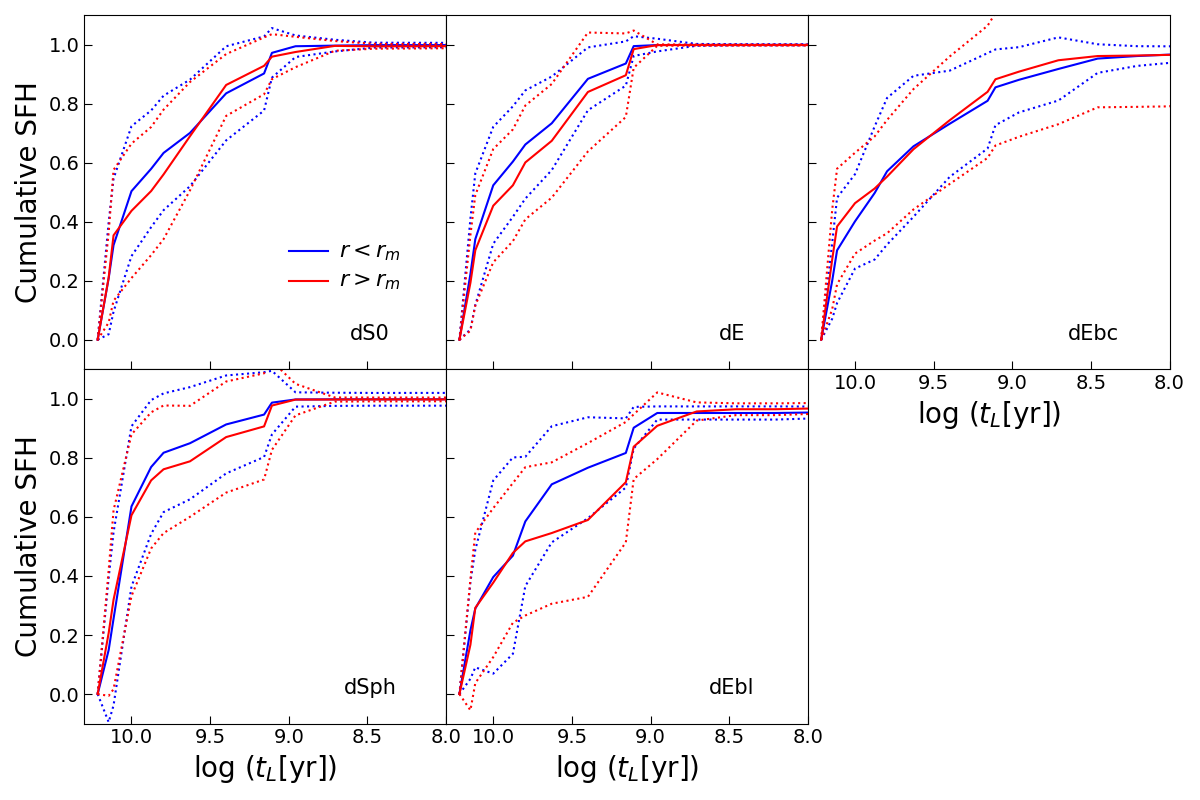}
			\caption{Cumulative star formation histories of early-type dwarf galaxies in the Virgo Cluster, divided into inner ($r < r_{m}$) and outer ($r > r_{m}$) regions, where $r_{m}$ is the mean clustercentric distance of the Virgo member galaxies. Solid lines represent the mean cSFHs and dotted lines indicate the $1\sigma$ envelope.}
			
			\label{fig13}
		\end{figure*}

Figure~\ref{fig13} presents the cSFHs of early-type dwarf galaxies in the Virgo Cluster, divided by clustercentric distance into an inner region ($r < r_{m}$) and an outer region ($r > r_{m}$). The cSFHs are shown separately for each morphological subtype: dS0, dE, dE$_{\mathbf{bc}}$, dSph and dE$_{\mathbf{bl}}$. As illustrated in Figure~\ref{fig13}, the initial phase of star formation, marked by an early starburst, exhibits little difference between the inner and outer regions across all subtypes. This similarity is especially notable in dSph galaxies, whose cSFHs remain nearly identical in both regions up to log~$t_L \approx 10$. Following the initial burst, however, the cSFHs of galaxies in the inner region increase more rapidly than those in the outer region, particularly between log~$t_L \approx 10$ and log~$t_L \approx 9$, indicating more sustained star formation in the inner cluster environment. An exception is the dS0 population, which shows virtually no difference between the inner and outer regions after log~$t_L \approx 9$. 

In general, the differences in cSFHs between the inner and outer regions of the Virgo Cluster are not substantial for most morphological types. However, dE$_{\mathbf{bl}}$ galaxies display a notable divergence between log~$t_L = 9.8$ and log~$t_L = 9.2$, despite exhibiting similar star formation activity during the earliest epoch. Although this difference appears prominent, it remains within the 1$\sigma$ confidence interval and is therefore not statistically significant, as indicated by the Kolmogorov–Smirnov test result of $p = 0.71$. This high $p$-value largely reflects the small number of dE$_{\mathbf{bl}}$ galaxies in the Virgo Cluster—only 11 in total, with just three residing in the inner region. Nonetheless, if the divergence between log~$t_L = 9.8$ and log~$t_L = 9.2$ reflects a genuine feature, it may be attributed to extended star formation in dE$_{\mathbf{bl}}$ galaxies located in the cluster outskirts, possibly sustained by accretion of cold gas from the surrounding environment or by fallback of gas from their own haloes. The presence of three dE$_{\mathbf{bl}}$ galaxies near the cluster centre is particularly intriguing, as the hot intracluster medium prevalent in the inner Virgo Cluster is expected to efficiently remove cold gas, thereby suppressing prolonged star formation.

To investigate this further, we examined the redshift and spatial information of the three inner-region dE$_{\mathbf{bl}}$ galaxies. Our analysis suggests that two of them are probable interlopers based on their redshift deviations, while the remaining one appears to be a bona fide Virgo Cluster member. If this is the case, the presence of a dE$_{\mathbf{bl}}$ galaxy near the cluster core may indicate a recently accreted system that has not yet experienced the full environmental effects of the cluster.

\section{Discussion}

\subsection{Pre-enrichment and Metal-free Accretion}

As shown in Figure~\ref{fig5}, the majority of stars formed during the first phase of star formation in early-type dwarf galaxies exhibit metallicities around $Z = 0.0004$. This relatively elevated value suggests that the interstellar medium (ISM) from which these stars formed was already enriched with heavy elements. Given that primordial nucleosynthesis in the Big Bang produced only hydrogen, helium, and trace amounts of lithium, such enrichment must have occurred prior to the onset of star formation in these systems.
In the absence of pre-enrichment, the first stars would be expected to have metallicities near zero ($Z \lesssim 0.0001$). The fact that stars in the first phase already display metallicities as high as $Z = 0.0004$ therefore provides compelling evidence for the early chemical enrichment of the ISM, most likely driven by massive Population~III stars in preceding cosmic structures.

Several observational and theoretical studies support the existence of pre-enrichment in both the ISM and the intergalactic medium (IGM). One of the most direct lines of evidence is the observed absence of extremely low metallicities—below $Z \approx 10^{-3}Z_{\odot}$—across a range of astrophysical environments \citep{cow98, hel06}. This metallicity floor has been identified in systems as diverse as Local Group dwarf spheroidal galaxies (e.g. \citealt{hel06}), damped Ly$\alpha$ absorbers \citep{wol98}, intervening IGM traced by quasar sightlines \citep[e.g.][]{cow98}, and the halo stars of the Milky Way.

The conspicuous absence of extremely metal-poor stars (e.g. $Z = 0.0001$) among the oldest stellar populations in most dSph galaxies reinforces the view that their natal gas was already enriched. This pre-enrichment raised the metallicity of early stellar populations to $Z \sim 0.0004$, in agreement with predictions from cosmological simulations \citep{tas12, wis12, jeon15}. These studies show that the metals produced by Population III supernovae rapidly enriched both the ISM of their host haloes and the surrounding IGM, establishing a baseline metallicity prior to the onset of subsequent star formation.

Signatures of pre-enrichment are evident in the star formation histories of dS0 galaxies \citep{ann24}, as well as in dSph and dE galaxies \citep{seo23}. In particular, the relatively high metallicity of $Z = 0.0004$ observed in the first stellar populations of early-type dwarfs strongly suggests that in-situ enrichment occurred during the epoch of reionisation. This early chemical enrichment may have been facilitated by an inhomogeneous gas surface density, leading to localized high-density pockets of pristine gas, ideal sites for the formation of Population~III stars \citep{tas12, pal14}. These stars would have rapidly synthesized and dispersed heavy elements into their surroundings, thereby raising the metallicity of the interstellar medium before the onset of subsequent star formation.

However, external enrichment may also have played a role. Gas clouds could have been enriched by Pop~III stars in nearby minihaloes or subhaloes, which were later accreted onto or merged with the host halo. This scenario introduces additional complexity to the enrichment history, suggesting that both internal and external mechanisms contributed to establishing the metallicity floor observed in the earliest stellar populations of early-type dwarf galaxies.

As discussed above, the star formation history of a galaxy is strongly influenced by its environment, and consequently, the evolution of its heavy-element content is also expected to vary significantly between galaxies in low-density regions (e.g. the field) and those in dense environments such as the Virgo Cluster. The top-left and bottom-right panels of Figure~\ref{fig6} show the stellar mass fractions as a function of metallicity for field and Virgo Cluster galaxies, respectively. By examining the distribution of stellar populations across metallicity bins, this figure illustrates the environmental dependence of chemical evolution.

In field galaxies, a small fraction of stars with extremely low metallicity ($Z = 0.0001$) formed during the first phase of star formation. In contrast, such metal-poor stars are virtually absent in Virgo Cluster galaxies. Even beyond the first phase, the metallicities of stars formed up to 10~Gyr ago in field galaxies remain relatively low, indicating a slower pace of chemical enrichment. This pattern is not observed in the Virgo Cluster, where the early stellar populations already exhibit significantly higher metallicities.

These trends suggest that pre-enrichment plays a substantially greater role in dense environments such as the Virgo Cluster, where early star formation occurred in gas already enriched by previous stellar generations. In contrast, galaxies in low-density regions like the field experienced weaker pre-enrichment, resulting in a more extended chemical evolution. As shown in Figure~\ref{fig5}, extremely metal-poor stellar populations formed during the first phase are predominantly found in dE$_{\mathbf{bc}}$ and dE$_{\mathbf{bl}}$ galaxies. Field galaxies exhibiting such characteristics can thus plausibly be associated with these morphological types.

Another notable feature in Figure~\ref{fig5} is the emergence of extremely metal-poor stars ($Z = 0.0001$) around a lookback time of $\sim$3.2~Gyr, particularly in dE$_{\mathbf{bc}}$ and dE$_{\mathbf{bl}}$ galaxies. This population is clearly visible in the field but absent in the Virgo Cluster. In field environments (see top-left panel of Figure~\ref{fig6}), the majority of stars formed during this episode exhibit $Z = 0.0001$, accounting for approximately 50~per~cent of the stellar mass formed at that time. The remainder consists of stars with intermediate metallicities ($Z \approx 0.004$ and $Z \approx 0.008$), reflecting gradual chemical enrichment following the initial star formation phase. The remarkably low metallicity of this population stands in sharp contrast to that of stars formed at the same epoch in the Virgo Cluster, where chemical enrichment was already more advanced and the inflow of metal-poor gas, responsible for the field's metal-poor star formation, was largely suppressed.

This second-phase accretion of pristine or nearly metal-free gas appears to be a phenomenon primarily associated with low-density environments. As the third phase of star formation progressed, such infall diminished, and subsequent star formation occurred from material that had undergone gradual chemical enrichment. Consequently, in field galaxies, stars formed during the later stages of this phase are predominantly characterized by $Z = 0.0004$. In contrast, stars formed during the same epoch in the Virgo Cluster exhibit much higher metallicities, typically around $Z = 0.05$. These findings highlight the role of environmental conditions in regulating gas accretion and chemical enrichment: late-time inflow of metal-poor gas is more feasible in the field, whereas it is largely suppressed in the dense intracluster medium of galaxy clusters like Virgo.

There is compelling evidence that star formation in early-type dwarf galaxies is significantly suppressed, primarily due to stellar feedback processes, particularly those associated with supernova explosions. These mechanisms were especially influential during the initial starburst phase, which occurred approximately 14~Gyr ago. The energy released by supernovae during this period likely heated the residual interstellar gas, thereby reducing the efficiency of subsequent star formation—particularly after the peak of star formation activity around 10~Gyr ago.

In addition, we observe so-called ‘gappy’ star formation histories in several dS0, dE, and dSph galaxies, consistent with findings by \citet{wri19}. These gappy SFHs are likely a consequence of supernova-driven feedback that is sufficiently strong to expel gas from the central regions into the halo, but not powerful enough to eject it into the intergalactic medium. The expelled gas remains gravitationally bound and gradually cools in the halo, eventually reaccreting onto the galaxy and triggering a subsequent episode of star formation. This cycle of gas expulsion and reaccretion provides a plausible explanation for the intermittent star formation observed in these systems, and underscores the critical role of internal feedback in shaping the long-term evolution of early-type dwarf galaxies.

\subsection{Nature versus Nurture}

The question of which physical processes primarily govern galaxy evolution: internal properties ('nature') or external environmental effects ('nurture') has long been a central topic in astrophysics. Galaxy evolution may be broadly divided into dynamical and chemical components: While the structure of a galaxy is shaped primarily by its dynamical history, its star formation history is more closely related to chemical evolution. This connection arises because, although hydrogen and helium were produced during Big Bang nucleosynthesis, all heavier elements (metals) were synthesized in stars and subsequently dispersed into the interstellar medium via stellar winds and supernova explosions. These metals enrich the gas reservoir and contribute to the formation of successive stellar generations, leading to a gradual increase in heavy-element content over time.

Figure~\ref{fig5} demonstrates that the rate and pattern of chemical enrichment vary with the detailed morphology of early-type dwarf galaxies. Despite this variation, all types exhibit a general trend of increasing metallicity over time. However, as shown in Figure~\ref{fig6}, this enrichment is also modulated by environment: the degree of chemical evolution differs markedly between field galaxies and those in dense environments such as the Virgo Cluster. These environmental differences partly reflect variations in the morphological composition of galaxies across environments. Since morphology itself is shaped by both internal processes and external influences, the interplay between galaxy type and environment plays a critical role in driving divergent evolutionary pathways.

As shown in Figure~\ref{fig6}, early-type dwarf galaxies in the field experience slow and extended chemical enrichment, whereas those in the Virgo Cluster undergo more rapid metal enrichment. The effect of pre-enrichment is considerably stronger in Virgo Cluster galaxies, where even the earliest stellar populations lack nearly metal-free stars. By contrast, field galaxies retain a small but distinct population of extremely metal-poor stars ($Z = 0.0001$), indicative of weaker early pre-enrichment.

An even more striking difference emerges during the third phase of star formation. In Virgo Cluster galaxies, this phase exhibits a sharp peak around 2.5~Gyr ago, whereas in field galaxies, it manifests as a broader and more prolonged episode spanning from 2.5 to 1~Gyr ago. This divergence is primarily driven by the accretion of metal-free gas in field galaxies during this phase. The infalling pristine gas mixes with residual material from earlier episodes, leading to the formation of metal-poor ($Z = 0.0004$) and extremely metal-poor ($Z = 0.0001$) stars. In contrast, galaxies in the cluster environment appear to lack such an inflow, resulting in uninterrupted enrichment and a more rapid chemical evolution during the third phase.

These differences suggest that the presence or absence of low-metallicity gas inflow plays a critical role in shaping the chemical enrichment history of a galaxy. From the perspective of the long-standing ‘nature versus nurture’ debate, our findings indicate that chemical evolution is more strongly governed by nurture—that is, by external environmental conditions. However, the picture is not entirely straightforward. Morphological type correlates with local galaxy density \citep{dre80}, implying that internal properties (nature) and environmental effects (nurture) are deeply intertwined. The disentanglement of their respective contributions to chemical evolution therefore remains a complex and nuanced challenge.

\begin{figure}
	\centering
	\includegraphics[width=0.5\textwidth]{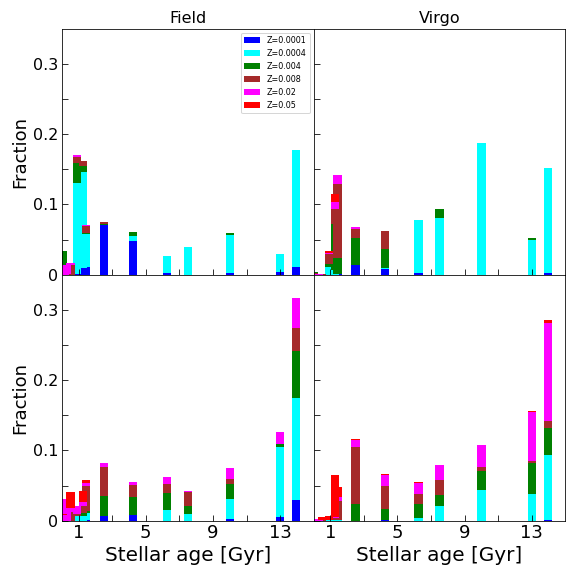}
	\caption{Stellar mass fractions of dS0, dE$_{\bold{bc}}$ and dSph galaxies in the field (left panels) and in the Virgo Cluster (right panels). Contributions from different metallicities are color-coded as indicated in the legend.}
	
    \label{fig14}	
\end{figure}

Figure~\ref{fig14} illustrates the combined effects of stellar mass and environment on the star formation histories of early-type dwarf galaxies. The left and right columns correspond to galaxies in the field and the Virgo Cluster, respectively, enabling direct environmental comparison. Within each column, the upper and lower panels represent low-mass and high-mass galaxies, respectively, allowing an assessment of mass dependence. The two mass bins are defined relative to the median stellar mass and its standard error: low-mass galaxies lie below the median minus one standard error ($\text{median} - \sigma$), and high-mass galaxies lie above the median plus one standard error ($\text{median} + \sigma$). 

Three key trends emerge from this figure, underscoring the interplay between stellar mass and environment in shaping SFHs. These features are discussed in the following.

First, Figure~\ref{fig14} reveals a strong dependence of star formation activity on stellar mass, with a more pronounced environmental effect in low-mass galaxies. In field galaxies, the stellar mass formed during the first phase of star formation is less than half that formed during the third phase, indicating more extended or delayed star formation in low-mass systems under low-density conditions. By contrast, low-mass galaxies in the Virgo Cluster show a more balanced SFH, with active star formation during the second phase—an episode largely absent in their field counterparts. Among high-mass galaxies, however, the SFH trend is largely independent of environment, with dominant early star formation during the first phase. This suggests that high-mass early-type dwarfs, regardless of environment, experienced intense and concentrated early star formation, likely due to deeper potential wells and more efficient gas retention.

Second, metal enrichment exhibits a strong dependence on both the stellar mass and the environment. In the low-mass group, stars formed during the first phase are predominantly metal-poor ($Z = 0.0004$), irrespective of environment. In contrast, high-mass galaxies display a broader range of stellar metallicities during the same epoch, from $Z = 0.0004$ to $Z = 0.02$. Notably, this mass- and environment-dependent chemical enrichment is already apparent in the earliest stellar populations formed during the initial starburst. The fraction of high-metallicity stars ($Z = 0.02$) produced in this phase varies significantly—ranging from approximately 10~per~cent in field galaxies to nearly 20~per~cent in Virgo Cluster galaxies—highlighting the earlier onset and greater efficiency of enrichment in dense environments.

Metal enrichment during the third phase of star formation also reveals distinct dependencies on both mass and environment. In field galaxies, low-mass systems continue to form primarily metal-poor stars until as recently as 1~Gyr ago, whereas high-mass galaxies in the same environment form stars with a broader metallicity range and relatively few metal-poor stars. In the Virgo Cluster, the difference in metallicity distribution between low- and high-mass galaxies is less pronounced during this phase—with the exception of a substantial population of extremely metal-rich stars ($Z = 0.05$) formed in high-mass systems after 1.5~Gyr ago. These trends suggest that chemical enrichment progresses most slowly in low-mass field galaxies and most rapidly in high-mass cluster galaxies, emphasizing the combined influence of both internal (mass-driven) processes and external (environmental) factors on the pace of chemical evolution.

Lastly, the accretion of metal-free gas appears to play a decisive role in shaping the third phase of star formation. This pristine gas mixes with the chemically enriched interstellar medium produced during earlier episodes, fueling a renewed burst of star formation. In low-mass field galaxies, this accretion is particularly prominent in dE$_{\mathbf{bl}}$ galaxies, although dE$_{\mathbf{bc}}$ galaxies also experience a non-negligible inflow of metal-free gas (Figure~\ref{fig5}). This accretion-driven star formation contributes more than 65~per~cent of the stellar mass formed during the third phase in low-mass field galaxies.

Importantly, this extended star formation episode in field galaxies—regardless of stellar mass—is largely sustained by continued gas inflow. The chemical evolution of these systems is strongly influenced by the metallicity of the accreted material. Notably, low-mass field galaxies did not form metal-rich stars until $\sim$1~Gyr ago, underscoring the suppressive effect of pristine gas on early enrichment. In contrast, high-mass field galaxies are less affected by metal-free accretion: although gas inflow continues until recent times, these systems still produce metal-rich stars during the third phase, suggesting that their deeper potential wells and higher star formation efficiencies mitigate the diluting effect of pristine gas.

These results underscore the decisive role of galaxy mass in regulating chemical enrichment in early-type dwarf galaxies. In this context, nature, specifically internal properties such as stellar mass, emerges as the dominant factor determining both the efficiency and extent of metal enrichment, whereas environmental effects (nurture) primarily modulate gas availability and the duration of star formation.

A closer examination of SFHs in the Virgo Cluster reveals that the peak epoch of star formation in low-mass galaxies occurs slightly later than in their high-mass counterparts. This delay is likely driven by the accretion of a metal-free gas, which is more prominent in low-mass systems and acts to prolong star formation. High-mass Virgo galaxies, in contrast, undergo rapid chemical evolution, largely because of the absence of such accretion. Consequently, the majority of stars formed by 1.5~Gyr ago in these systems already possess high metallicities, with $Z = 0.05$ being the dominant heavy-element abundance.

It is particularly noteworthy that a non-negligible fraction of stars with $Z = 0.05$ continues to form even in recent epochs. This suggests that the gas fueling ongoing star formation in these high-mass cluster galaxies is primarily recycled, rather than accreted from external metal-poor reservoirs. The absence of pristine gas inflow in these environments reinforces the importance of internal gas cycling in sustaining late-time star formation in massive galaxies residing in dense environments. These findings underscore that in the absence of external gas supply, internal processes such as feedback and recycling become increasingly dominant in shaping the late-stage evolution of early-type dwarf galaxies.

\section{Conclusions}

We have investigated the star formation histories (SFHs) of 983 early-type dwarf galaxies, classified into five morphological subtypes—dS0, dE, dE$_{\mathbf{bc}}$, dSph, and dE$_{\mathbf{bl}}$—across six environmental categories ranging from the field to rich clusters such as Ursa Major (UMa) and Virgo. By applying full spectral fitting using the {\sc starlight} code to SDSS spectra, we derived detailed SFHs and chemical enrichment patterns, enabling a comprehensive exploration of how morphology and environment shape galaxy evolution.

The SFHs of early-type dwarf galaxies generally exhibit two major episodes of star formation. The initial phase, comprising one or two bursts around 14 and 10 Gyr ago, contributes significantly to the stellar mass budget. Most stars formed during this epoch are metal-poor ($Z = 0.0004$), with a small fraction at $Z = 0.0001$, implying substantial pre-enrichment even at early times.

Morphology is found to be the dominant factor shaping the global SFH structure. Red early-type dwarfs (dS0, dE, dSph) undergo rapid early star formation and early quenching, with $\tau_{90}$ typically earlier than 2 Gyr ago. Among them, dSph galaxies show the most truncated SFHs, quenching about 3.3 Gyr ago. In contrast, blue early-type dwarfs (dE$_{\mathbf{bc}}$ and dE$_{\mathbf{bl}}$) form stars over more extended timescales and exhibit delayed chemical evolution, often including extremely metal-poor stellar populations.

While morphology governs the broad structure of the SFH, environmental effects are clearly evident in shaping the timing and duration of later star formation, especially during the second period. Galaxies in high-density environments such as Virgo tend to quench earlier and lack late-epoch metal-poor stars. However, the extended SFHs of blue early-type dwarfs persist across environments, indicating that these systems sustain star formation through continued gas accretion.

Cumulative SFHs confirm these trends: dSph galaxies formed most of their stellar mass before 6.3 Gyr ago, while dE and dS0 galaxies typically quenched 1 Gyr ago. In contrast, blue subtypes often continue to form stars until recent times, particularly in low-density environments. Within Virgo, galaxies in the cluster outskirts show more prolonged star formation than those in the core, reflecting differences in mass and local conditions.

Stellar mass also plays a critical role in modulating environmental effects. At low masses, Virgo galaxies quench much earlier than their field counterparts, whereas high-mass galaxies display similar SFHs across environments. This suggests that environmental quenching is more effective in shallow potential wells, highlighting the interplay between external pressure and internal gravitational depth.

In the third phase of star formation, accretion of pristine gas fuels renewed activity, especially in low-mass field galaxies. This process dilutes the interstellar medium and delays chemical enrichment, resulting in the prolonged formation of metal-poor stars. In contrast, high-mass galaxies form metal-rich stars even during this phase, suggesting that deeper potentials mitigate dilution effects. In Virgo, high-mass galaxies lack evidence of pristine gas inflow and undergo rapid chemical evolution, with $Z = 0.05$ stars forming as early as 1.5 Gyr ago. A non-negligible fraction of such stars continues to form even in recent times, likely from recycled internal gas.

In summary, the evolutionary pathways of early-type dwarf galaxies are determined by a complex interplay between morphology, stellar mass, and environment. Although morphology controls the overall shape of the SFH, environmental conditions and mass jointly influence quenching timescales, gas accretion, and the pace of metal enrichment. Our results underscore the diversity of early-type dwarfs and demonstrate that they follow multiple, intersecting evolutionary tracks shaped by both nature and nurture.

\section*{Acknowledgements}

H. B. Ann thanks to Dr. Cid Fernandes for the source of STARLIGHT  which is essential for the present study. M. J. is supported by the National Research Foundation (NRF) grants funded by the Korean government (MSIT) (No.2021R1A2C109491713, No. 2022M3K3A1093827).



\section*{DATA AVAILABILITY} 

The original data underlying this article are available in the SDSS DR7.
Additional data are available upon request.  


\bsp    
\label{lastpage}

\end{document}